\documentclass[10pt, conference, letterpaper]{IEEEtran}

\RequirePackage{graphics}
\usepackage{epsfig}
\usepackage{amsmath}
\usepackage{amsfonts}
\usepackage{graphicx,subfigure}
\usepackage{amssymb}
\usepackage{algorithmic}
\usepackage{algorithm,float}
\usepackage{color}
\usepackage{framed}
\usepackage{xcolor}
\usepackage{lipsum}

\newcommand{\tabincell}[2]{\begin{tabular}{@{}#1@{}}#2\end{tabular}}
\def\overbracket#1{\mathop{\vbox{\ialign{##\crcr\noalign{\kern3\p@}
\downbracketfill\crcr\noalign{\kern3\p@\nointerlineskip}
$\hfil\displaystyle{#1}\hfil$\crcr}}}\limits}
\def\underbracket#1{\mathop{\vtop{\ialign{##\crcr
$\hfil\displaystyle{#1}\hfil$\crcr\noalign{\kern3\p@\nointerlineskip}
\upbracketfill\crcr\noalign{\kern3\p@}}}}\limits}

\makeatletter
\newenvironment{breakablealgorithm}
  {
   \begin{center}
     \refstepcounter{algorithm}
     \hrule height.8pt depth0pt \kern2pt
     \renewcommand{\caption}[2][\relax]{
       {\raggedright\textbf{\ALG@name~\thealgorithm} ##2\par}%
       \ifx\relax##1\relax 
         \addcontentsline{loa}{algorithm}{\protect\numberline{\thealgorithm}##2}%
       \else 
         \addcontentsline{loa}{algorithm}{\protect\numberline{\thealgorithm}##1}%
       \fi
       \kern2pt\hrule\kern2pt
     }
  }{
     \kern2pt\hrule\relax
   \end{center}
  }
\makeatother

\usepackage{cite}

\ifCLASSINFOpdf
\else
\fi

\begin{document}

\title{Covert Wireless Communications \\ with Active Eavesdropper on AWGN Channels}


\author{\IEEEauthorblockN{
Zhihong Liu\IEEEauthorrefmark{1}, Jiajia Liu\IEEEauthorrefmark{1}\IEEEauthorrefmark{5}, Yong
Zeng\IEEEauthorrefmark{1}, Jianfeng Ma\IEEEauthorrefmark{1}, and Qiping Huang \IEEEauthorrefmark{2}}
\IEEEauthorblockA{\IEEEauthorrefmark{1}School of Cyber Engineering, Xidian University, Xi'an, China}
\IEEEauthorblockA{\IEEEauthorrefmark{2}School of Telecommunication Engineering, Xidian University, Xi'an,
China} \IEEEauthorblockA{\IEEEauthorrefmark{5}E-mail: liujiajia@xidian.edu.cn} }


\maketitle

\begin{abstract}
Covert wireless communication can prevent an adversary from knowing the existence of user's transmission, thus
provide stronger security protection. In AWGN channels, a square root law was obtained and the result shows
that Alice can reliably and covertly transmit $\mathcal{O}(\sqrt{n})$ bits to Bob in $n$ channel uses in the
presence of a passive eavesdropper (Willie). However, existing work presupposes that Willie is static and only
samples the channels at a fixed place. If Willie can dynamically adjust the testing distance between him and Alice
according to his sampling values, his detection probability of error can be reduced significantly via a trend test.
We found that, if Alice has no prior knowledge about Willie, she cannot hide her transmission behavior in the
presence of an active Willie, and the square root law does not hold in this situation. We then proposed a novel
countermeasure to deal with the active Willie. Through randomized transmission scheduling, Willie cannot detect
Alice's transmission attempts if Alice can set her transmission probability below a threshold. Additionally, we
systematically evaluated the security properties of covert communications in a dense wireless network, and
proposed a density-based routing scheme to deal with multi-hop covert communication in a wireless network. As
the network grows denser, Willie's uncertainty increases, and finally resulting in a ``shadow'' network to Willie.
\end{abstract}


\IEEEpeerreviewmaketitle

\section{Introduction}
Wireless networks are changing the way we interact with the world around us. Billions of small and smart wireless
nodes can communicate with each other and cooperate to fulfil sophisticated tasks. However, due to the inherent
openness of wireless channels, the widespread of wireless networks and development of pervasive computing not
only opens up exciting opportunities for economic growth, but also opens the door to a variety of new security
threats.

Traditional network security methods based on cryptography can not solve all security problems. If a wireless
node wishes to talk to other without being detected by an eavesdropper, encryption is not enough
\cite{Hiding_Information}. Even a message is encrypted, the pattern of network traffic can reveal some sensitive
information. Additionally, if the adversary cannot ascertain Alice's transmission behavior, Alice's communication is
unbreakable even if the adversary has unlimited computing and storage resources and can mount powerful
quantum attacks \cite{Computer21}. On other occasions, users hope to protect their source location privacy
\cite{panda_hunter}, they also need to prevent the adversary from detecting their transmission attempts.

Covert communication has a long history. It is always related with steganography \cite{Steganography} which
conceals information in covertext objects, such as images or software binary code. While steganography requires
some forms of content as cover, the network covert channel requires network protocols as carrier
\cite{covert_channel_1}\cite{covert_channel_2}. Another kind of covert communication is spread spectrum
\cite{Spread_Spectrum} which is used to protect wireless communication from jamming and eavesdropping. In
this paper, we consider another kind of physical-layer covert wireless communications that employs noise as the
cover to hide user's transmission attempts.

Consider the scenario where Alice would like to talk to Bob over a wireless channel in order to not being detected
by a warden Willie. Bash \emph{et al.} found  a square root law \cite{square_law} in AWGN channels: Alice can
only transmit $\mathcal{O}(\sqrt{n})$ bits reliably and covertly to Bob over $n$ uses of wireless channels. If
Willie does not know the time of transmission attempts of Alice, Alice can reliably transmit more bits to Bob with
a slotted AWGN channel \cite{time1}. In practice, Willie has measurement uncertainty about its noise level due to
the existence of SNR wall \cite{SNR}, then Alice can achieve an asymptotic privacy rate which approaches a
non-zero constant \cite{LDP1}\cite{Biao_He}.  In discrete memoryless channels (DMC), the privacy rate of
covert communication is found to scale like the square root of the blocklength \cite{Fundamental_Limits}. Bloch
\emph{et al.} \cite{Bloch} discussed the covert communication problem from a resolvability perspective, and
developed an alternative coding scheme to achieve the covertness.

In general, the covertness of communication is due to the existence of noise that Willie cannot accurately
distinguish Alice's signal from the background noise.  To improve the performance of covert communication,
interference or jamming can be leveraged as a useful security tool
\cite{Challenges}\cite{Artificial_noise_Goel}\cite{heartbeats}. In \cite{jammer1}, Sober \emph{et al.} added a
friendly ``jammer'' to wireless environment to help Alice for security objectives. Soltani \emph{et al.}
\cite{jammer2}\cite{DBLP:journals/corr/abs-1709-07096} considered a network scenario where there are
multiple ``friendly'' nodes that can generate jamming signals to hide the transmission attempts from multiple
adversaries. Liu \emph{et al.} \cite{The_Sound_and_the_Fury} and He \emph{et al.} \cite{Biao_he_2} studied
the covert wireless communication with the consideration of interference uncertainty. From the network
perspective, the communications are hidden in the noisy wireless networks.

For the methods discussed above, the eavesdropper Willie is assumed to be passive and static, which means that
Willie is placed in a fixed place, eavesdropping and judging Alice's behavior from his observations. However, an
active Willie can launch other sophisticated attacks. Willie is active does not mean he can interact with other
nodes involved. An active Willie is a passive eavesdropper who can dynamically adjust the distance between him
and Alice according to his sampling value to make more accurate test. At the beginning, Willie is far away from
Alice, gathering samples of the background noise, and employing a radiometer to detect Alice's behavior. If he
finds his observations look suspicious, Willie moves to a closer place for further detection. After having gathered
a number of samples at different places, Willie makes a decision on whether Alice is transmitting or not. We
found that, if Alice has no prior knowledge about Willie, she cannot hide her transmission behavior in the presence
of an active Willie in her vicinity, and the square root law does not hold in this situation. Willie can easily detect
Alice's transmission attempts via a trend test. To deal with the active Willie, we propose a novel countermeasure
to increase the detection difficulty of Willie, and  then present a density-based routing scheme for multi-hop
covert communication in a dense wireless network.

The primary contributions of this paper are summarized as follows.
\begin{enumerate}
  \item We introduce an active Willie for the first time and show that the square root law is no longer valid in
      the presence of the active Willie. Besides, other covert communication methods, such as interference or
      cooperative jamming,  have little effect on the covertness in the presence of the active Willie.
  \item To deal with the active Willie, we propose countermeasures to confuse Willie further. We show that,
      through a randomized transmission scheduling, Willie cannot detect Alice's transmission attempts for a
      certain significance value if Alice's transmission probability is set below a threshold.
  \item We further study the covert communication in dense wireless networks, and propose a density-based
      routing (DBR) to deal with multi-hop covert communications. We find that, as the network becomes more
      and more denser and complicated,  Willie's difficulty of detection is greatly increased, and nonuniform
      network is securer than a uniform network.
\end{enumerate}

The remainder of this paper is organized as follows. Section \ref{ch_2} describes the system model, In Section
\ref{ch_3}, we present the active Willie attack. The countermeasures to the active Willie are studied in Section
\ref{ch_4}, and the covert communication in  a dense wireless network is discussed in Section \ref{ch_5}. Finally,
Section \ref{ch_6} concludes the paper and discusses possible future research directions.

\section{System Model}\label{ch_2}
\subsection{Channel Model}
Consider a wireless communication scene where Alice (A) wishes to transmit messages to Bob (B). An
eavesdropper, or a warden Willie (W) is eavesdropping over wireless channels and trying to find whether or not
Alice is transmitting. We adopt the wireless channel model similar to \cite{square_law}. Each node, legitimate node
or eavesdropper, is equipped with a single omnidirectional antenna. All wireless channels are assumed to suffer
from discrete-time AWGN with real-valued symbols, and the wireless channel is modeled by large-scale fading
with path loss exponent $\alpha$ ($\alpha > 2$).

Let the transmission power employed for Alice be $P_0$, and $s^{(A)}$ be the real-valued symbol Alice
transmitted which is a Gaussian random variable $\mathcal{N}(0,1)$. The receiver Bob observes the signal
$y^{(B)} = s^{(A)}+z^{(B)}$, where $z^{(B)}\sim \mathcal{N}(0, \sigma^2_{B,0})$ is the noise Bob experiences.
As to Willie, he observes the signal $y^{(W)} = s^{(A)}+z^{(W)}$, and $z^{(W)}$ is the noise Willie experiences
with $z^{(W)}\sim \mathcal{N}(0, \sigma^2_{W,0})$. Suppose Bob and Willie experience the same background
noise power, i.e., $\sigma^2_{B,0} = \sigma^2_{W,0}=\sigma^2_0$. Then, the signal seen by Bob and Willie can
be represented as follows,
\begin{eqnarray}
  y^{(B)}  &\equiv & \sqrt{\frac{P_0}{d_{A,B}^{\alpha}}} \cdot s^{(A)}+z^{(B)}\sim \mathcal{N}(0,\sigma^2_{B}) \label{eq_1}\\
  y^{(W)}  &\equiv & \sqrt{\frac{P_0}{d_{A,W}^{\alpha}}} \cdot s^{(A)}+z^{(W)}\sim \mathcal{N}(0,\sigma^2_{W}) \label{eq_2}
\end{eqnarray}
and
\begin{eqnarray}
\sigma^2_{B}&=&\frac{P_0}{d_{A,B}^{\alpha}}+\sigma^2_0 \\
\sigma^2_{W}&=&\frac{P_0}{d_{A,W}^{\alpha}}+\sigma^2_0  \label{eq_4_4}
\end{eqnarray}
where $d_{A,B}$ and $d_{A,W}$ are the Euclidean distances between Alice and Bob, Alice and Willie,
respectively.

\subsection{Active Willie}
In \cite{square_law} and \cite{DBLP:journals/corr/abs-1709-07096}, the eavesdropper Willie is assumed to be
passive and static, which means that Willie is placed in a fixed place, eavesdropping and judging Alice's behavior
from his samples $y^{(W)}_1, y^{(W)}_2, \cdots, y^{(W)}_n$ with each sample $y^{(W)}_i\sim
\mathcal{N}(0,\sigma^2_{W})$. Based on the sampling vector $\mathbf{y}=(y_1^{(W)}, \cdots, y_n^{(W)})$,
Willie makes a decision on whether the received signal is noise or Alice's signal plus noise. Willie employs a
radiometer as his detector, and does the following statistic test
\begin{equation}
    T(\mathbf{y})=\frac{1}{n}\mathbf{y}^H\mathbf{y}=\frac{1}{n}\sum^n_{k=1}y_k^{(W)}*y_k^{(W)}>\gamma
\end{equation}
where $\gamma$ denotes Willie's detection threshold and $n$ is the number of samples.

\begin{figure}
\centering \epsfig{file=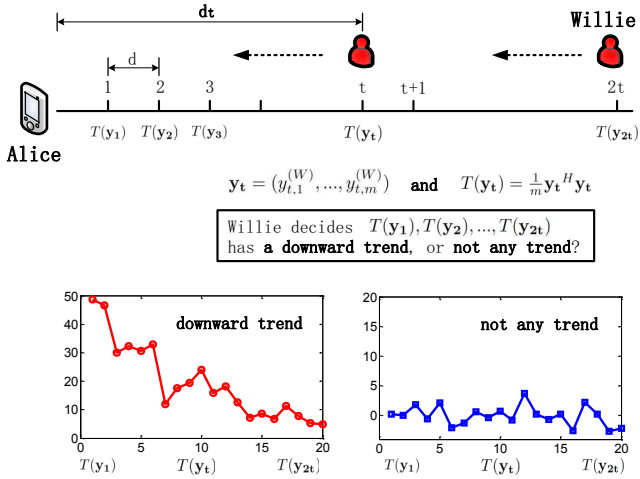, height=2.5in}
\caption{Covert wireless communication in the presence of an active Willie who leverages a trend analysis to detect Alice's transmission attempts.}\label{framework}
\end{figure}

The system framework with an active Willie is depicted in Fig. \ref{framework}. Willie can move to several places
to gather more samples for further detection.

In the Fig. \ref{framework},  Willie detects Alice's behavior  at $2t$ different locations (each location is $d$
meters apart). At each location he gathers $m$ samples. For example, at $t$-th location, Willie's samples can be
presented as a vector
\begin{equation}
    \mathbf{y_t}=(y_{t,1}^{(W)}, y_{t,2}^{(W)}, \cdots, y_{t,m}^{(W)})
\end{equation}
where each sample $y^{(W)}_{t,m}\sim \mathcal{N}(0,\sigma^2_{W})$.

The average signal power at $t$-th location can be calculated as follows
\begin{equation}
    T(\mathbf{y_t})=\frac{1}{m}\mathbf{y_t}^H\mathbf{y_t}.
\end{equation}
Therefore Willie will have a signal power vector $\mathbf{T}$, consisting of $2t$ values at different locations
\begin{equation}
    \mathbf{T}=(T(\mathbf{y_1}), T(\mathbf{y_2}), \cdots, T(\mathbf{y_t}), \cdots, T(\mathbf{y_{2t}}))
\end{equation}
Then Willie decides whether $(T(\mathbf{y_1}), T(\mathbf{y_2}), \cdots, T(\mathbf{y_{2t}}))$ has a
downward trend or not any trend. If the trend analysis shows a downward trend for given significance level
$\beta$, Willie can ascertain that Alice is transmitting  with probability $1-\beta$.

\subsection{Hypothesis Testing}
To find whether Alice is transmitting or not, Willie has to distinguish between the following two
hypotheses,
\begin{eqnarray}
\mathbf{H_0}&:& \text{there is not any trend in vector $\mathbf{T}$;} \\
                      & & y^{(W)}  =  z^{(W)}  \nonumber\\
\mathbf{H_1}&:& \text{there is a downward trend in vector $\mathbf{T}$.}\\
                      & & y^{(W)}  =  \sqrt{\frac{P_0}{d_{A,W}^\alpha}}\cdot s^{(A)}+ z^{(W)} \nonumber
\end{eqnarray}

Given the vector $\mathbf{T}$, Willie can leverage the Cox-Stuart test \cite{cox} to detect the presence of
trend. The idea of the Cox-Stuart test is based on the comparison of the first and the second half of the
samples. If there is a downward trend, the observations in the second half of the samples should be smaller than
in the first half. If they are greater, the presence of an upward trend is suspected. If there is not any trend one
should expect only small differences between the first and the second half of the samples due to randomness.

The Cox-Stuart test is a sign test applied to the sample of non-zero differences. To perform a trend analysis on
$\mathbf{T}$, the sample of differences is to be calculated as follows
\begin{eqnarray}
  \Delta_1 &=& T(\mathbf{y_1})-T(\mathbf{y_{t+1}}) \nonumber\\
  \Delta_2 &=& T(\mathbf{y_2})-T(\mathbf{y_{t+2}}) \nonumber\\
     & &  \cdot\cdot\cdot\cdot\cdot\cdot \nonumber\\
  \Delta_t &=& T(\mathbf{y_t})-T(\mathbf{y_{2t}}) \nonumber
\end{eqnarray}
Let $sgn(\Delta_i)=1$ for $\Delta_i<0$, and suppose the sample of negative differences by $\Delta_1,
\Delta_2, ..., \Delta_k$, then the test statistic of the Cox-Stuart test on the vector $\mathbf{T}$ is
\begin{equation}
\mathbf{T}_{\Delta<0}=\sum^k_{i=1}sgn(\Delta_i)
\end{equation}

Given a significance level $\beta$ and the binomial distribution $\mathbf{b}\sim b(t,0.5)$, we can reject the
hypothesis $\mathbf{H_0}$ and accept the alternative hypothesis $\mathbf{H_1}$ if
$\mathbf{T}_{\Delta<0}<\mathbf{b}(\beta)$ which means a downward trend is found with probability larger
than $1-\beta$, where $\mathbf{b}(\beta)$ is the  quantile of the binomial distribution $\mathbf{b}$.
According to the central limit theorem, if $t$ is large enough ($t>20$), an approximation
$\mathbf{b}(\beta)=1/2[t+\sqrt{t}\cdot\Phi^{-1}(\beta)]$ can be applied, where $\Phi^{-1}(\beta)$ is the
$\beta$-quantile function of the standard normal distribution. Therefore, if
\begin{equation}\label{qq}
    \mathbf{T}_{\Delta<0}<\frac{1}{2}[t+\sqrt{t}\cdot\Phi^{-1}(\beta)],
\end{equation}
Willie can ascertain that Alice is transmitting with probability larger than $1-\beta$ for the significance level
$\beta$ of test.

The parameters and notation used in this paper are illustrated in Table \ref{tab_1}.

\begin{table}
\caption{parameters and notation}
\begin{tabular}{|l|l|}
  \hline
  $\mathbf{Symbol}$ & $\mathbf{Meaning}$ \\
  \hline
  $P_0$ & Transmit power of Alice \\
  \hline
  $t$ & \tabincell{l}{Number of differences in Cox-Stuart test. \\ There are $2t$ sampling points.}  \\
  \hline
    $\alpha$ & Path loss exponent \\
  \hline
  $m$ & Number of samples in a sampling location \\
  \hline
  $P_i$ & Willie's received power at $i$-th sampling location \\
  \hline
  $d_i$ & Distance between Alice and Willie's $i$-th location \\
    \hline
  $d$ & Spacing between sampling points \\
    \hline
  $s^{(A)}$ & Alice's signal  \\
  \hline
  $y^{(B)}$, $y^{(W)}$ & Signal (Bob, Willie) observes  \\
  \hline
  $z^{(B)}$, $z^{(W)}$ & (Bob's, Willie's) background noise \\
  \hline
  $\sigma_{B,0}^2, ~\sigma_{W,0}^2$ & Power of noise (Bob, Willie) observes  \\
  \hline
  $\mathbf{y_t}$ & Willie's samples at $t$-th location \\
  \hline
  $T(\mathbf{y_t})$ & Signal power at $t$-th location \\
  \hline
  $\mathbf{T}$ & Signal power vector \\
  \hline
  $\lambda$ & Density of the network \\
  \hline
  $\chi^2_t(m)$ & \tabincell{l}{Chi-square distribution \\ with m degrees of freedom (at $t$-th location) } \\
  \hline
  $\mathcal{N}(\mu,\sigma^2)$ & \tabincell{l}{Gaussian distribution with mean $\mu$ and  variance $\sigma^2$} \\
  \hline
  $\mathbf{T}_{\Delta<0}$ &  Test statistic of the Cox-Stuart test \\
  \hline
  $\beta$ &  Significance level of testing \\
  \hline
  $\mathbf{E}[X]$ & Mean of random variable $X$ \\
   \hline
   $\mathbf{Var}[X]$ & Variance of random variable $X$ \\
   \hline
   $\Phi^{-1}(\beta)$ & $\beta$-quantile function of $\mathcal{N}(0, 1)$ \\
   \hline
\end{tabular}\label{tab_1}
\end{table}

\section{Active Willie Attack} \label{ch_3}
This section discusses the covert wireless communication in the presence of active Willie. Bash \emph{et al.} found
a square root law in AWGN channels \cite{square_law}, which implies that Alice can transmit
$\mathcal{O}(\sqrt{n})$ bits reliably and covertly over $n$ uses of channels. However, if an active Willie is
placed in the vicinity of Alice, Alice cannot conceal her transmitting behavior. Willie can detect Alice's transmission
behavior with arbitrarily low probability of error.

As illustrated in Fig. \ref{framework},  when Alice is transmitting, Willie detects Alice's behavior  at $2t$
different locations. At each location he gathers $m$ samples and calculates the signal power at this location. At
$t$-th location (with the distance $d_t$ between Alice and Willie), Willie's samples are a vector
\begin{equation}
    \mathbf{y_t}=(y_{t,1}^{(W)}, y_{t,2}^{(W)}, \cdots, y_{t,m}^{(W)})
\end{equation}
with $y^{(W)}_{t,i}\sim \mathcal{N}(0,\sigma^2_{W})$  and the signal power at this location is
\begin{equation}
    T(\mathbf{y_t})=\frac{1}{m}\mathbf{y_t}^H\mathbf{y_t}\sim \frac{P_t+\sigma_0^2}{m}\chi_t^2(m)
\end{equation}
where $P_t=P_0d_t^{-\alpha}$ and $\chi_t^2(m)$ is the chi-squared distribution with $m$ degrees of freedom.

\begin{figure} \centering
\subfigure[Alice is silent]{ \epsfig{file=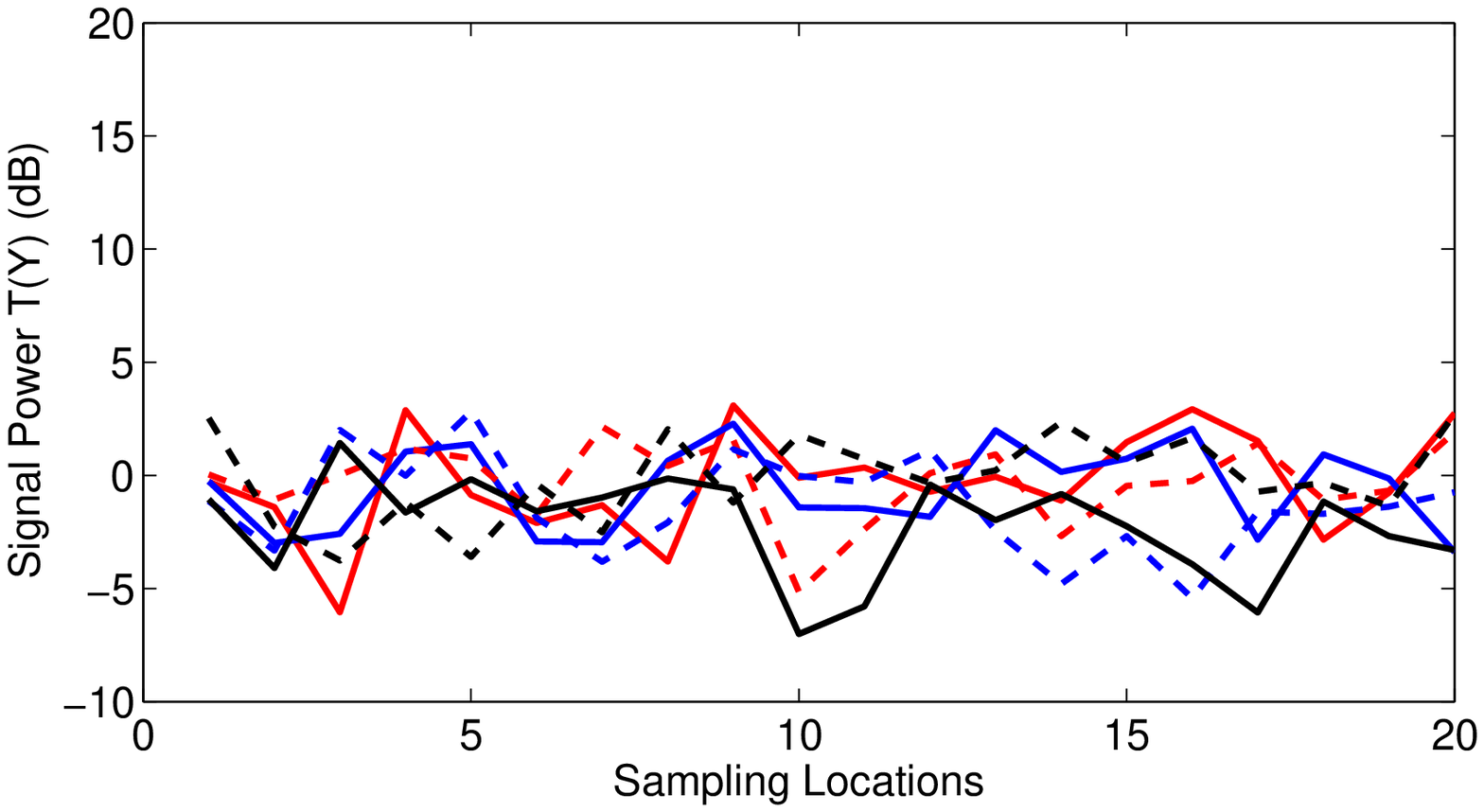, height=1.7in }}
\subfigure[Alice is transmitting]{ \epsfig{file=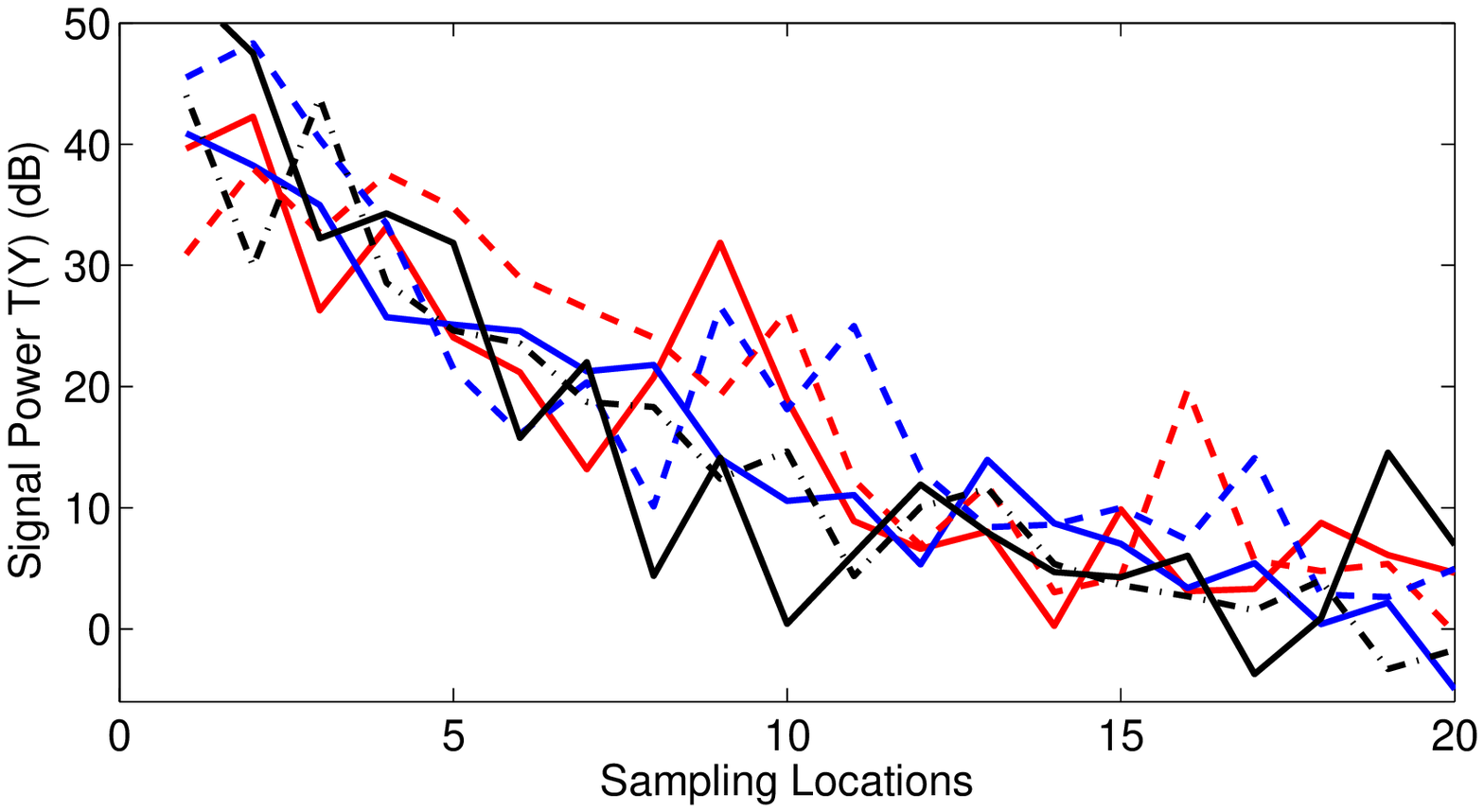, height=1.7in} }
\caption{The signal power at different locations when (a) Alice is silent, or (b) Alice is transmitting.
Here a bounded path loss law is used, $l(x)=\frac{1}{1+\parallel x\parallel^\alpha}$. The transmit power $P_0$ of Alice is 30dB,
links experience unit mean Rayleigh fading, and $\alpha=3$. The spacing between sampling locations $d=0.5$m. }
\label{trend01}
\end{figure}

Fig. \ref{trend01} shows examples of the signal power at different locations when Alice is transmitting or not.
We can find that, even if the channel experiences fading, the downward trend of the signal power is obvious when
Alice is transmitting.

Next we discuss the method that Willie utilizes to detect transmission attempts. With $2t$ values
$T(\mathbf{y_i})$  at different locations in his hand, Willie decides whether $(T(\mathbf{y_1}),
T(\mathbf{y_2}), \cdots, T(\mathbf{y_{2t}}))$ has a downward trend or not. If Alice is transmitting, the
probability that the difference $\Delta_i=T(\mathbf{y_i})-T(\mathbf{y_{t+i}})<0$ can be estimated as follows,
\begin{eqnarray}
   \mathbb{P}\{\Delta_i<0\} &=& \mathbb{P}\{T(\mathbf{y_i})<T(\mathbf{y_{t+i}})\} \nonumber\\
   &=&  \mathbb{P}\biggl\{\frac{P_i+\sigma_0^2}{m}\chi_i^2(m)<\frac{P_{t+i}+\sigma_0^2}{m}\chi_{t+i}^2(m)\biggr\}\nonumber\\
   &=& \mathbb{P}\biggl\{\frac{\chi_{t+i}^2(m)}{\chi_i^2(m)}>\frac{P_i+\sigma_0^2}{P_{t+i}+\sigma_0^2}\biggr\} \nonumber\\
   &\leq& \mathbb{P}\biggl\{\biggl|\frac{\chi_{t+i}^2(m)}{\chi_i^2(m)}\biggr|>\frac{P_{i}+\sigma_0^2}{P_{t+i}+\sigma_0^2}\biggr\}
\end{eqnarray}
where $P_i=P_0d_i^{-\alpha}$ and $P_{t+i}=P_0d_{t+i}^{-\alpha}$.

Given a random variable $\mathbf{X}$ and its PDF $f_{\mathbf{X}}(x)$, its second moment can be bounded as
\begin{eqnarray}
  \mathbf{E}[\mathbf{X}^2] &=& \int_{-\infty}^{\infty}x^2f_{\mathbf{X}}(x)\mathrm{d}x > \int_{|\mathbf{X}|\geq t}x^2f_{\mathbf{X}}(x)\mathrm{d}x \nonumber\\
   &\geq& t^2\int_{|\mathbf{X}|\geq t}f_{\mathbf{X}}(x)\mathrm{d}x =t^2\cdot\mathbb{P}\{|\mathbf{X}|\geq t\}
\end{eqnarray}
Hence
\begin{equation}\label{ss}
    \mathbb{P}\{|\mathbf{X}|\geq t\}\leq \frac{\mathbf{E}[\mathbf{X}^2]}{t^2}
\end{equation}

Because $\chi_i^2(m)$ and $\chi_{t+i}^2(m)$ are two independent chi-squared distributions with $m$ degrees
of freedom, then the random variable $\mathbf{Y}$ is
\begin{equation}\label{ww}
    \mathbf{Y}=\frac{\chi_{t+i}^2(m)}{\chi_i^2(m)}\sim F(m,m)
\end{equation}
where $F(m,m)$ is an F-distribution with two parameters $m$ and $m$, and its mean and variance are
\begin{equation}\label{fmm}
 \mathbf{E}[\mathbf{Y}]=\frac{m}{m-2}, ~~ \mathbf{Var}[\mathbf{Y}]=\frac{4(m-1)m}{(m-2)^2(m-4)} ~~(m>4)
\end{equation}
and when $m\rightarrow\infty$, $\mathbf{E}[\mathbf{Y}]=1$, $\mathbf{Var}[\mathbf{Y}]=0$.

If Willie can gather enough samples at each location, i.e., $m$ is large, according to (\ref{ss}) and (\ref{fmm}),
we have
\begin{eqnarray}\label{equation1}
   \mathbb{P}\{\Delta_i<0\} &\leq& \mathbb{P}\biggl\{|\mathbf{Y}|>\frac{P_{i}+\sigma_0^2}{P_{t+i}+\sigma_0^2}\biggr\} \nonumber\\
   &\leq& \frac{\mathbf{Var}[\mathbf{Y}]+\mathbf{E}[\mathbf{Y}]^2}{\bigl(\frac{P_{i}+\sigma_0^2}{P_{t+i}+\sigma_0^2}\bigr)^2} \nonumber\\
   &=& \biggl(\frac{P_{t+i}+\sigma_0^2}{P_{i}+\sigma_0^2}\biggr)^2
\end{eqnarray}

Therefore the number of negative  differences in $\Delta_1, \Delta_2, ..., \Delta_t$ can be estimated as
follows
\begin{equation}\label{df}
    \mathbf{T}_{\Delta<0}=\sum^t_{i=1}\mathbf{1}_{\{\Delta_i<0\}}\leq \sum^t_{i=1}\biggl(\frac{P_{t+i}+\sigma_0^2}{P_{i}+\sigma_0^2}\biggr)^2
\end{equation}
where $\mathbf{1}_{\{\Delta_i<0\}}$ is an indicator function, $\mathbf{1}_{\{\Delta_i<0\}}=1$ when
$\Delta_i<0$; otherwise $\mathbf{1}_{\{\Delta_i<0\}}=0$.

As to Willie, his received signal strength at $i$-th location is $P_i=P_0d_i^{-\alpha}=P_0(i\cdot d)^{-\alpha}$
which is a decreasing function of the distance $d_i$.  Suppose Willie knows the power level of noise. At first
Willie monitors the environment. When he detects the anomaly with $P_{i}\geq\sigma_0^2$, Willie then
approaches Alice to carry out more stringent testing.

According to the setting, we have
\begin{equation}
    P_1>P_2>\cdots>P_{2t-1}>P_{2t}=\sigma_0^2
\end{equation}
and
\begin{eqnarray}
  P_1-P_2 &>& P_{t+1}-P_{t+2},  \nonumber\\
  P_2-P_3 &>& P_{t+2}-P_{t+3}, \nonumber\\
  \cdots &\cdots&  \cdots \nonumber\\
  \frac{P_1}{P_2} &>& \frac{P_{t+1}}{P_{t+2}}, \\
  \frac{P_2}{P_3} &>& \frac{P_{t+2}}{P_{t+3}}, \nonumber\\
    \cdots &\cdots&  \cdots \nonumber
\end{eqnarray}
Thus
\begin{equation}
    \frac{P_{t+1}+\sigma_0^2}{P_{1}+\sigma_0^2}<\frac{P_{t+2}+\sigma_0^2}{P_{2}+\sigma_0^2}<\cdots<\frac{P_{2t}+\sigma_0^2}{P_{t}+\sigma_0^2}
\end{equation}
Since $P_t=2^{\alpha}P_{2t}$ and $P_{2t}\geq\sigma_0^2$,
\begin{equation}
    \frac{P_{2t}+\sigma_0^2}{P_{t}+\sigma_0^2}=\frac{P_{2t}+\sigma_0^2}{2^{\alpha}P_{2t}+\sigma_0^2}=\frac{1+\frac{\sigma_0^2}{P_{2t}}}{2^{\alpha}+\frac{\sigma_0^2}{P_{2t}}}\leq\frac{2}{2^{\alpha}+1}
\end{equation}
According to Equ. (\ref{equation1}), the negative differences have
\begin{equation}\label{df}
    \mathbf{T}_{\Delta<0}\leq \sum^t_{i=1}\biggl(\frac{P_{t+i}+\sigma_0^2}{P_{i}+\sigma_0^2}\biggr)^2\leq t\cdot\biggl(\frac{2}{2^{\alpha}+1}\biggr)^2
\end{equation}
If $t$ is large enough, the following inequality holds
\begin{equation}
    \mathbf{T}_{\Delta<0}\leq t\cdot\biggl(\frac{2}{2^{\alpha}+1}\biggr)^2< \frac{1}{2}[t+\sqrt{t}\cdot\Phi^{-1}(\beta)].
\end{equation}

Therefore, given any small significance level $\beta>0$, if the number of differences $t$ satisfies
\begin{equation}
  t>\biggl(\frac{\Phi^{-1}(\beta)}{1-\frac{8}{(2^{\alpha}+1)^2}}\biggr)^2,
\end{equation}
Willie can distinguish between two hypotheses $\mathbf{H_0}$ and  $\mathbf{H_1}$ with probability
$1-\beta$.

\begin{figure}
\centering \epsfig{file=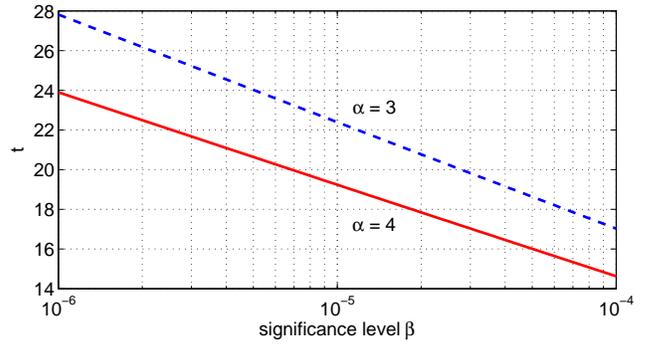, height=1.8in}
\caption{The significance level $\beta$ versus $t$ for the path loss exponent $\alpha=3$ and $\alpha=4$.}\label{t_beta}
\end{figure}

Fig. \ref{t_beta} shows the significance level $\beta$ versus $t$ for different path loss exponent.  Less
significance level $\beta$,  more sampling locations Willie should take to distinguish two hypotheses. If Alice is
transmitting, Willie can ascertain that Alice is transmitting with probability $1-\beta$ for any small $\beta$. This
may be a pessimistic result since it demonstrates that Alice cannot resist the attack of active Willie and the
square root law \cite{square_law} does not hold in this situation. If Willie can move to the vicinity of Alice and
have enough sampling locations, Alice is no longer able to hide her transmission attempts.

\section{Countermeasures to Active Willie} \label{ch_4}
The previous discussion shows that, Alice cannot hide her transmission behavior in the presence of an active Willie,
even if she can utilize other transmitters (or jammers) to increase the interference level of Willie, such as the
methods used in \cite{jammer1}\cite{jammer2}
\cite{DBLP:journals/corr/abs-1709-07096}\cite{The_Sound_and_the_Fury}. These methods can only raise the
noise level but not change the trend of the sampling value. Next we discuss two methods that can increase the
detection difficulty of the active Willie.

\subsection{Dynamic power adjustment}
If Alice has information about Willie, such as Willie's location, the simplest way she can take is decreasing her
transmission power when she finds out that Willie is in her close proximity. When Willie is very close, Alice simply
stops transmitting and keeps quiet until Willie leaves.

However,  Alice may be a small and simple IoT device who is not able to perceive the environmental information,
let alone knowing Willie's location. As to the active Willie, he is a passive eavesdropper who does not interact with
any of the parties involved, attempting to determine the transmission party. Therefore Willie cannot be easily
detected.

\begin{figure}
\centering \epsfig{file=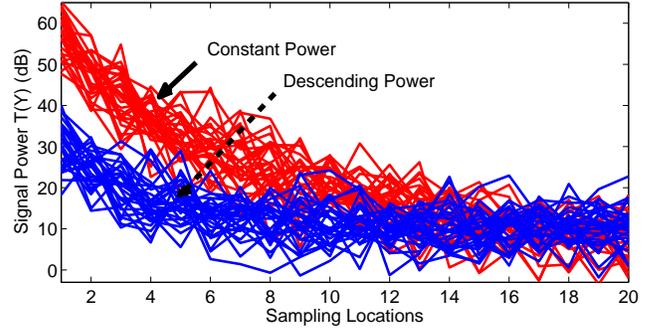, height=1.8in}
\caption{Constant transmission power versus descending transmission power. Here constant power is set to 33dB, and the transmission power is
descended from 36dB to 20dB.  The spacing between sampling locations $d=0.5$m.}\label{decrease_power}
\end{figure}

In practice, Alice can dynamically adjust her transmission power to make the decreasing tendency of her signal
power $T(\mathbf{Y})$ unclear to Willie's detector.  As depicted in Fig. \ref{decrease_power}, Alice chooses
the maximum transmission power $p_{max}=36$dB and the minimum transmission power $p_{min}=20$dB, and at
first transmits at the maximum transmission power, then decreases $\Delta=0.8$dB at each location.  From the
figure we find, if Alice transmits with decreasing power and Willie approaches Alice to take further samples, the
signal power of Willie's detector has a weaker growth trend than the constant transmission power Alice employed.
When Willie is far away from Alice, the signal power Willie sees has no certain trend, just like the background
noise. Only when Willie approaches Alice, a growth trend gradually increases and could be detected easily.

This approach can only be used in the occasion that Willie cannot sneak up on Alice very closely.  Besides, if Willie
gradually moves away from Alice and gathers the signal power in this procedure, he will see a significant
downward trend in his signal power, resulting in the exposure of Alice's transmission behavior.

\subsection{Randomized transmission scheduling}
\begin{figure} \centering
\subfigure[p=0.1]{ \epsfig{file=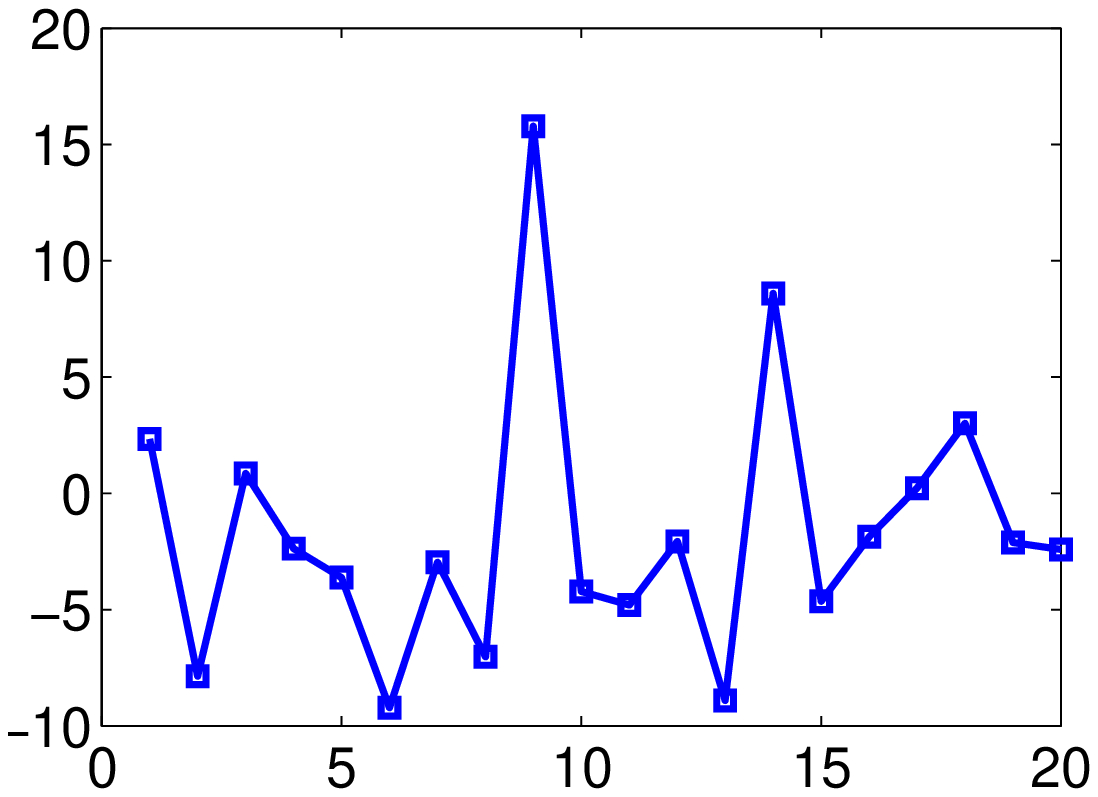, height=0.77in }}
\subfigure[p=0.5]{ \epsfig{file=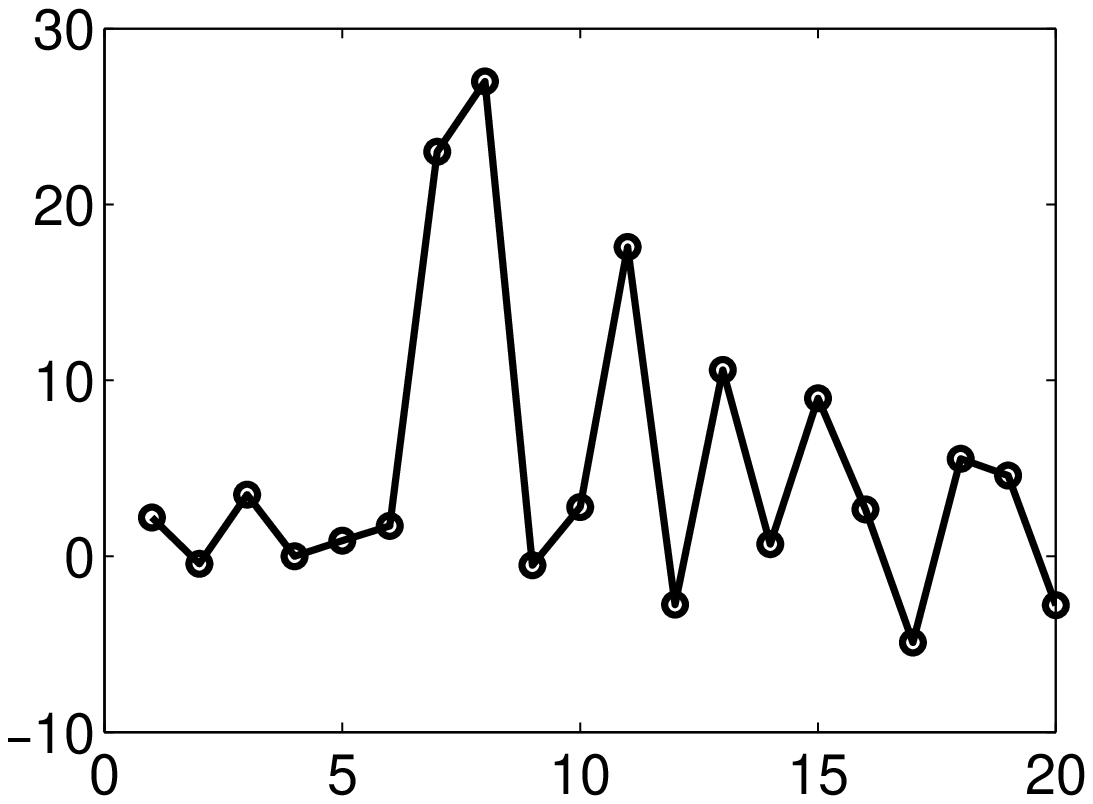, height=0.77in} }
\subfigure[p=0.9]{ \epsfig{file=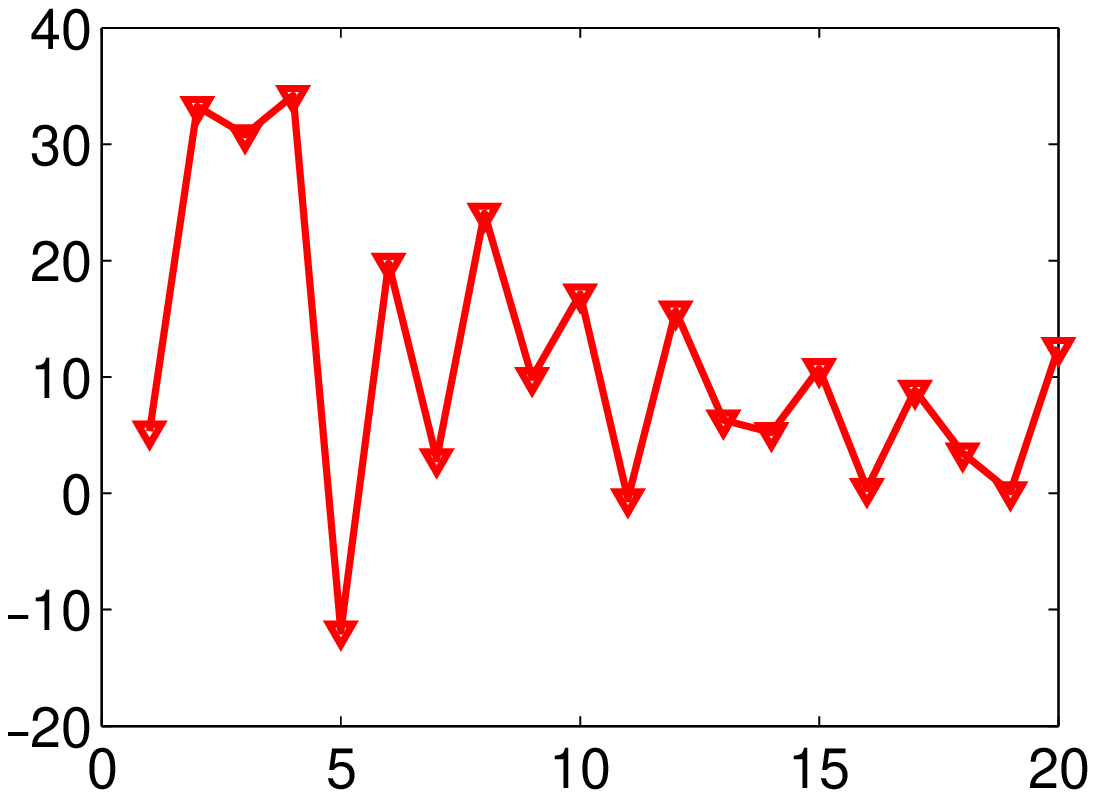, height=0.77in} }
\subfigure[p=0.1]{ \epsfig{file=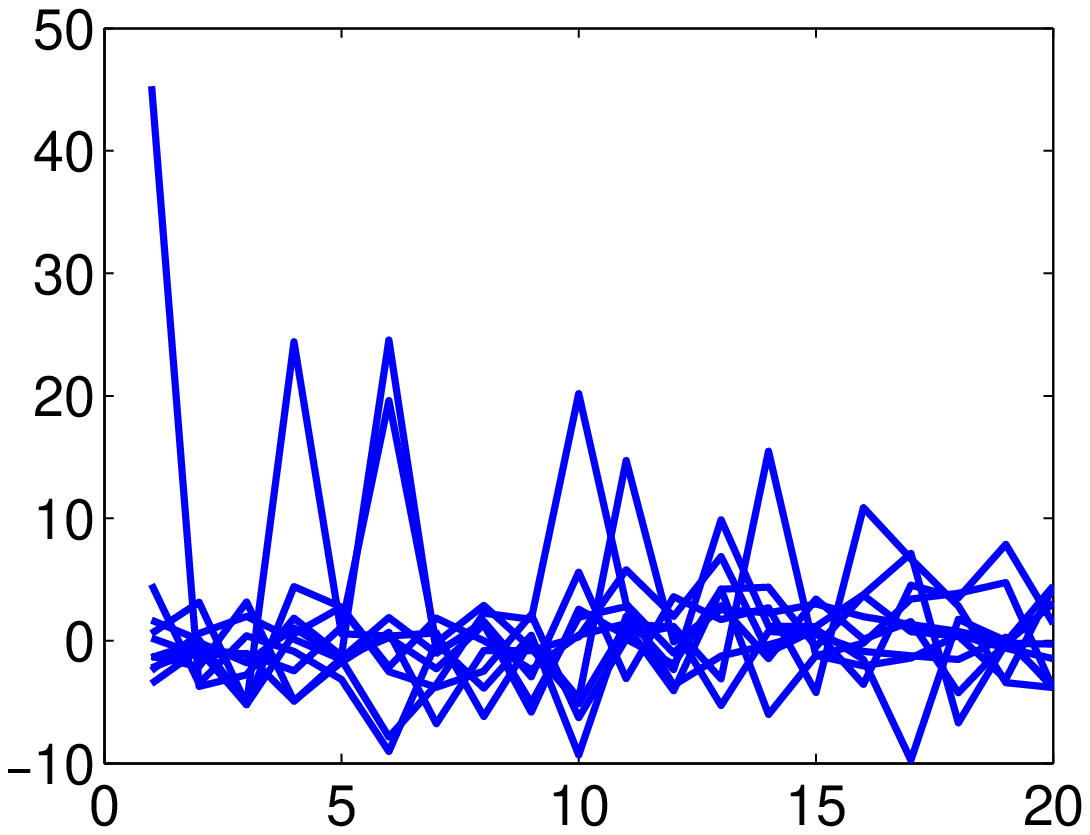, height=0.8in }}
\subfigure[p=0.5]{ \epsfig{file=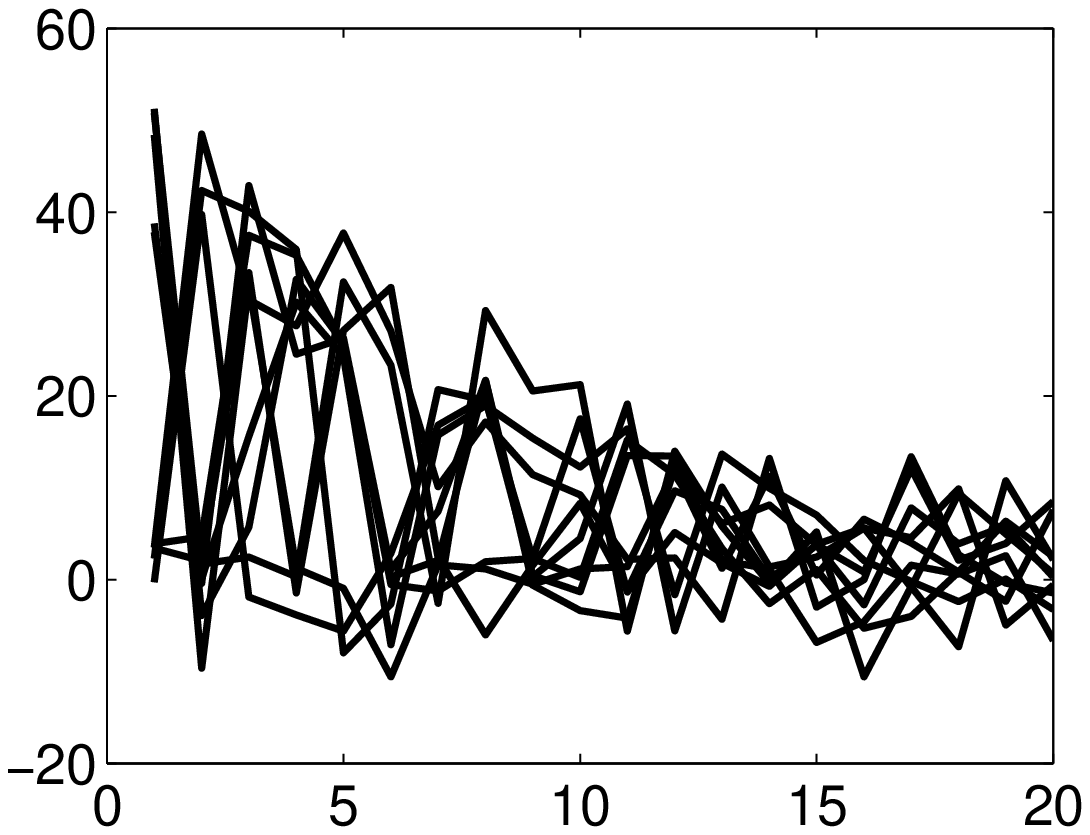, height=0.8in} }
\subfigure[p=0.9]{ \epsfig{file=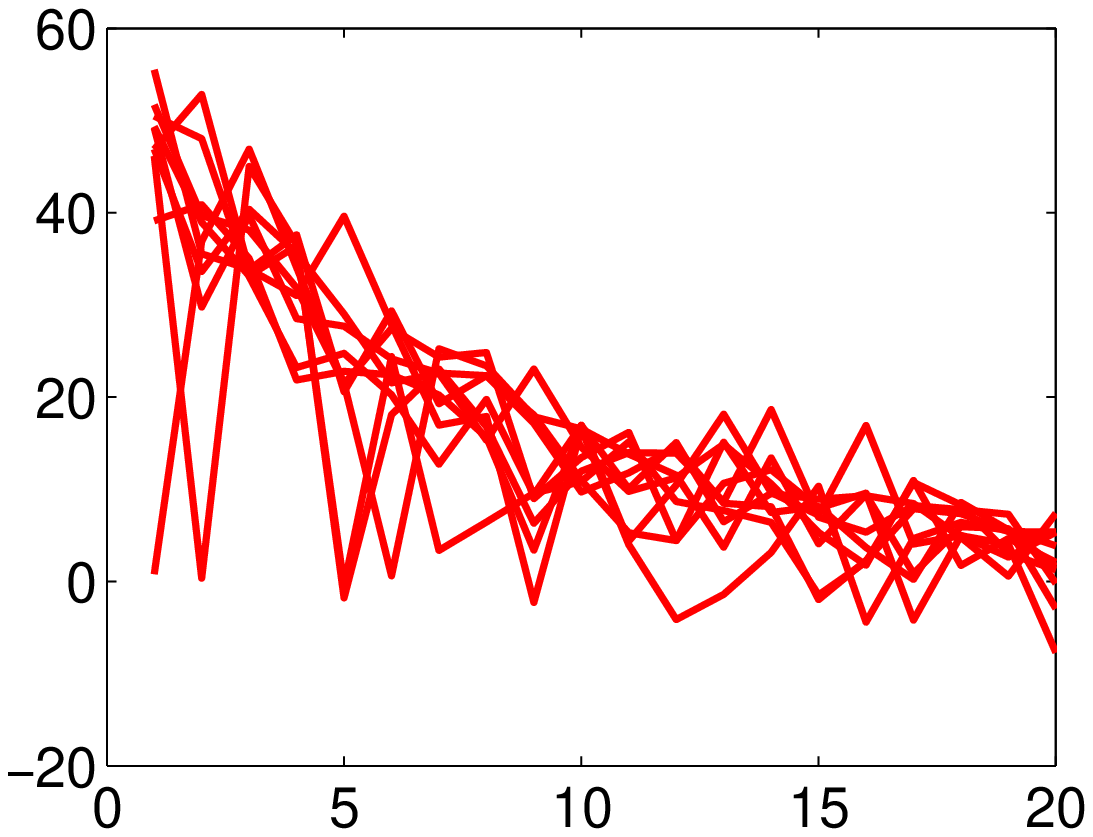, height=0.8in} }
\caption{The signal power at different  locations with the transmission probability (a) p=0.1, (b) p=0.5, (c) p=0.9,
(d) p=0.1 multi-testing, (e) p=0.5 multi-testing, and (f) p=0.9 multi-testing. }
\label{scheduling}
\end{figure}

In practice, to confuse Willie,  Alice's transmitted signal should resemble white noise. Alice should not generate
burst traffic, but transforming a bulk message into a slow network traffic with transmission and silence
alternatively. She can divide the time into slots, then put the message into small packets. After that, Alice sends a
packet in a time slot with a transmission probability $p$, or keeps silence with the probability $1-p$, and so on.
Fig. \ref{scheduling} illustrates the examples of Willie's sampling signal power $T(\mathbf{y_1}), \cdots,
T(\mathbf{y_{2t}})$  in the case that Alice's transmission probability $p$ is set to 0.1, 0.5, and 0.9. Clearly,
when Alice's transmission probability $p$ decreases, the downward tendency of the signal power is lessening,
Willie's uncertainty increases.

Next we evaluate the effect of the transmission probability on the covertness of communications. Suppose Alice
divides the time into slots, and transmits at a slot with the transmission probability $p$. As to Willie, for each time
slot, he samples the channel $m$ times at a fixed location, then at the next slot he moves to a closer location to
sample the channel, and so on. The signal power Willie obtained at $i$-th sampling location can be represented as
follows
\begin{equation}
T(\mathbf{y_i}) =
\begin{cases}
\frac{\sigma_0^2}{m}\chi_i^2(m)& \mathbf{X}=0 \\
\frac{P_i+\sigma_0^2}{m}\chi_i^2(m)& \mathbf{X}=1
\end{cases}
\end{equation}
where $\mathbf{X}$ is a random variable, $\mathbf{X}=1$ if Alice is transmitting, $\mathbf{X}=0$ if Alice is
silent, and the transmission probability $\mathbb{P}\{\mathbf{X}=1\}=p$.

Given the transmitting probability $p$, the probability that the difference
$\Delta_i=T(\mathbf{y_i})-T(\mathbf{y_{t+i}})<0$ can be estimated as follows
\begin{eqnarray}\label{aaqq}
   \mathbb{P}\{\Delta_i<0\} &=& \mathbb{P}\{T(\mathbf{y_i})<T(\mathbf{y_{t+i}})\} \nonumber\\
   &=& (1-p)^2\cdot\mathbb{P}\biggl\{\frac{\chi_{i}^2(m)}{\chi_{i+t}^2(m)}<1\biggr\} \nonumber\\
   & & + p(1-p)\cdot\mathbb{P}\biggl\{\frac{\chi_{i}^2(m)}{\chi_{i+t}^2(m)}<\frac{P_{i+t}+\sigma_0^2}{\sigma_0^2}\biggr\} \nonumber\\
   & & + p(1-p)\cdot\mathbb{P}\biggl\{\frac{\chi_{i}^2(m)}{\chi_{i+t}^2(m)}<\frac{\sigma_0^2}{P_{i}+\sigma_0^2}\biggr\} \nonumber\\
   & & + p^2\cdot \mathbb{P}\biggl\{\frac{\chi_{i}^2(m)}{\chi_{i+t}^2(m)}<\frac{P_{i+t}+\sigma_0^2}{P_{i}+\sigma_0^2}\biggr\}
\end{eqnarray}
where $P_i=P_0d_i^{-\alpha}$ and $P_{t+i}=P_0d_{t+i}^{-\alpha}$.

Since $\mathbb{P}\{F(m,m)<1\}=\frac{1}{2}$, thus when $p$ is small enough, we can approximate Equ.
(\ref{aaqq}) as follows
\begin{eqnarray}
\mathbb{P}\{\Delta_i<0\} &\rightarrow& (1-p)^2\cdot\mathbb{P}\biggl\{\frac{\chi_{i}^2(m)}{\chi_{i+t}^2(m)}<1\biggr\} \nonumber \\
                                       &=&(1-p)^2\cdot\mathbb{P}\{F(m,m)<1\} \nonumber \\
                                       &=& \frac{1}{2}(1-p)^2
\end{eqnarray}
and the test statistic of the Cox-Stuart test is
\begin{equation}\label{qq}
    \mathbf{T}_{\Delta<0}=\sum_{i=1}^t\mathbb{P}\{\Delta_i<0\}\rightarrow\frac{1}{2}(1-p)^2\cdot t
\end{equation}

Therefore, for any small significance value $\beta$, Alice can find a proper transmission probability $p$ which
satisfies
\begin{equation}\label{qq}
    \mathbf{T}_{\Delta<0}=\frac{1}{2}(1-p)^2\cdot t> \frac{1}{2}[t+\sqrt{t}\cdot\Phi^{-1}(\beta)]
\end{equation}
This means that, when the transmission probability $p$ is set to
\begin{equation}
    p<1-\sqrt{1+\frac{\Phi^{-1}(\beta)}{\sqrt{t}}}, ~~ t>[\Phi^{-1}(\beta)]^2
\end{equation}
then Willie cannot detect Alice's transmission behavior for a certain significance value $\beta$.

\begin{figure}
\centering \epsfig{file=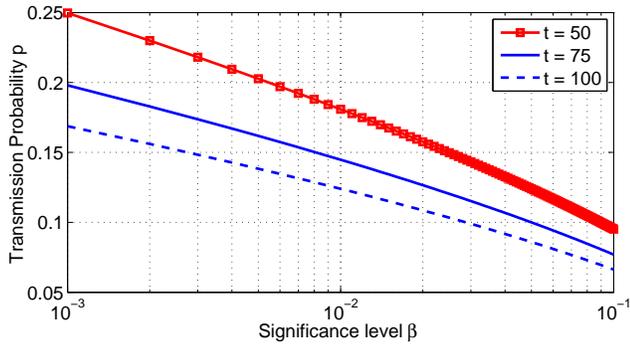, height=1.8in}
\caption{The significance level $\beta$ versus the transmission probability $p$ for the number of differences in Cox-Stuart test,
$t=50$, $t=75$, and $t=100$.}\label{p_beta}
\end{figure}

Fig. \ref{p_beta} depicts the significance level $\beta$ versus the transmission probability $p$ for different
parameter $t$ in the Cox-Stuart test. Larger significance level $\beta$ will result in lower transmission
probability $p$, and more differences in trend test $t$ will increase Willie's detecting ability, which implies that
Alice should decrease her transmission probability.

The randomized transmission scheduling is a practical way for Alice to decrease the probability of being detected.
Although it may increases the transmission latency, Willie's uncertainty increases as well.

\section{Covert Wireless Communication in Dense Networks} \label{ch_5}
In practice, to detect the transmission attempt of Alice, Willie should approach Alice as close as possible, and
ensure that there is no other node located closer to Alice than him. Otherwise, Willie cannot determine which one
is the actual transmitter.


\begin{figure}
\centering \epsfig{file=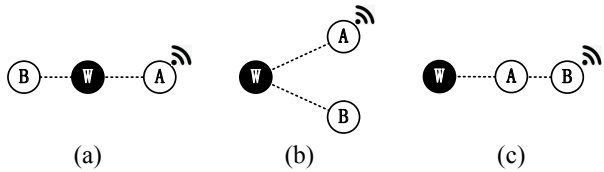, height=0.9in} \caption{Examples of covert wireless communication in network with static Willie.}\label{three_case}
\end{figure}

In a dense wireless network, Bob and Willie not only experience noise, but also interference from other
transmitters simultaneously. In this scenario, it is difficult for Willie to detect a certain  transmitter tangibly. Fig.
\ref{three_case}  illustrates the dilemma Wille faced in covert wireless communication with multiple potential
transmitters. As shown in Fig. \ref{three_case}(a) and (b), if Willie (W) finds his observations look suspicious, he
knows for certain that some nodes are transmitting, but he cannot determine whether Alice (A) or Bob (B) is
transmitting. Even in the case of Fig. \ref{three_case} (c), Willie cannot determine with confidence that Alice (A),
not Bob (B) is transmitting, since the received signal strength of Willie is determined by the randomness of Alice's
signal and the fading of wireless channels. Therefore it is difficult to be predicted.

%
\begin{figure}
\centering \epsfig{file=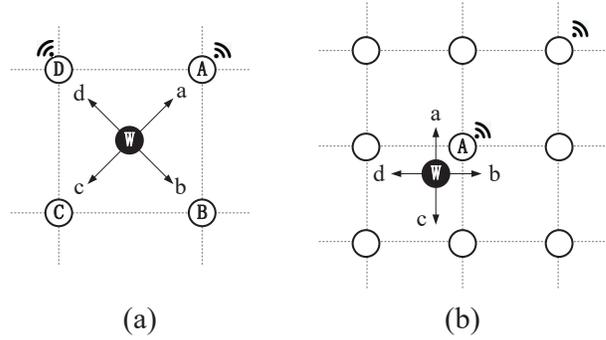, height=1.8in} \caption{Covert wireless communication in a dense network.}\label{network2}
\end{figure}

For a static and passive Willie, to discriminate the actual transmitter from the other is an almost impossible task,
provided that there is no obvious radio fingerprinting of transmitters can be exploited \cite{radio_fingerprinting}.
And what's worse is, Willie will be bewildered by a dense wireless network with a large number of nodes.  As
depicted in Fig. \ref{network2} (a), Willie has detected suspicious signals, but among nodes A, B, C, and D who are
the real transmitters is not clear. To check whether A is a transmitter, Willie could move to A along the direction
$a$ and sample at different locations of this path. If he can find a upward trend in his samples via Cox-Stuart
test, he could ascertain that A is transmitting. If a weak downward trend can be found, Wille can be sure that A
is not a transmitter, but C's suspicion increases. Willie may move along the direction $c$ to carry out more
accurate testing. If a upward trend is found, then C may be the transmitter. However, if there is no trend found,
the transmitters may be B or D.  Then Willie could move along direction $b$ or $d$ to find the actual signal
source. This is a slight simplification, and more complicated scenarios are possible. In the case depicted in Fig.
\ref{network2} (b), Willie should move along all four directions, sampling, and testing the existence of any trend
to distinguish Alice among her many neighbors, potential transmitters.

\begin{figure}
\centering \epsfig{file=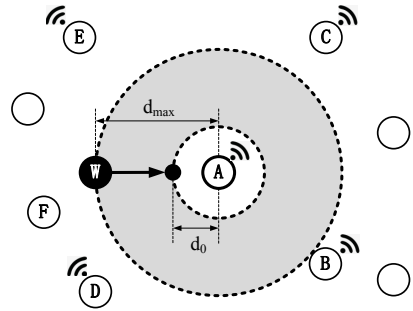, height=1.6in}
\caption{The detection region of the active Willie (the gray area). Here $d_0$ is the minimum distance and $d_{max}$ is the maximum distance that Willie can take.}\label{fig_condition2}
\end{figure}

In a dense wireless network Willie cannot always be able to find out who is transmitting. As depicted in Fig.
\ref{fig_condition2}, the gray area is the detection region of Willie where there is no other potential transmitters
and $d_{max}$ is the maximum distance that can be used by Willie to take his trend statistical test. However
Willie cannot get too close to Alice, $d_0$ is the minimum distance between them. In a wireless network, some
wireless nodes are probably placed on towers, trees, or buildings, Willie cannot get close enough as he wishes. If
Willie leverages the Cox-Stuart trend test to detect Alice's transmission behavior, the following conditions should
be satisfied:
\begin{itemize}
  \item No other node in the detection region of Willie.
  \item In the detection region, Willie should have enough space for testing, i.e., Willie can take as many
      sampling points as possible, and the spacing of points is not too small.
\end{itemize}

\begin{figure}
\centering \epsfig{file=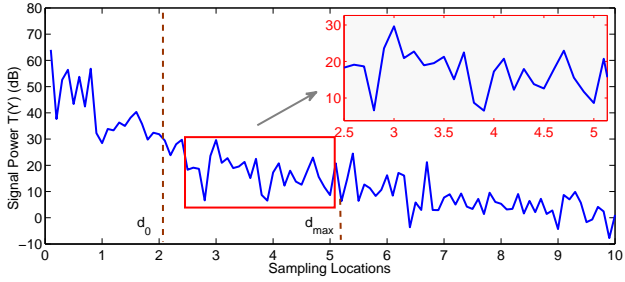, height=1.5in}
\caption{The signal power at different sampling locations.}\label{network_wave}
\end{figure}

As  illustrated in Fig. \ref{network_wave}, if Willie can only get samples in the distance interval $(d_0,
d_{max})=(2.5,5)$m, it is difficult to discover a downward trend in the signal power vector he obtains in a short
testing interval.

Suppose $d_{A,W}$ be the distance between Alice and Willie, then the probability that there is no nodes inside
the disk region with Alice as the centre and $d_{A,W}$ as a radius can be estimated as follows
\begin{equation}
    \mathbb{P}\{N=0\} = \exp(-\pi\lambda d^2_{A,W})
\end{equation}
where $\lambda$ is the density of the network.

Let $\mathbb{P}\{N=0\} >1-\epsilon$ for small $\epsilon>0$, we have
\begin{equation}
d_{A,W}<\sqrt{\frac{1}{\pi\lambda}\ln\frac{1}{1-\epsilon}}
\end{equation}

In a random graph $G(n,p)$, if $c=pn>1$, then the largest component of $G(n,p)$ has $\Theta(n)$ vertices, and
the second-largest component has at most $\frac{16c}{(c-1)^2}\log n$ vertices a.a.s. If $c=pn=1$, the largest
component has $\Theta(n^{\frac{2}{3}})$ vertices \cite{Haenggi_PPP}.

Using the above considerations, we have
\begin{equation}
c=pn=\frac{\lambda\pi d^2_{A,W}}{n}\cdot n=\lambda\pi d^2_{A,W}>1
\end{equation}
then if the density of the wireless network satisfies the following condition
\begin{equation}
\lambda>\frac{1}{\pi d^2_{A,W}}
\end{equation}
the wireless network will become a shadow network to Willie since nodes  are so close that he cannot distinguish
between them in a very narrow space.

\begin{figure*} \centering
\subfigure[]{ \epsfig{file=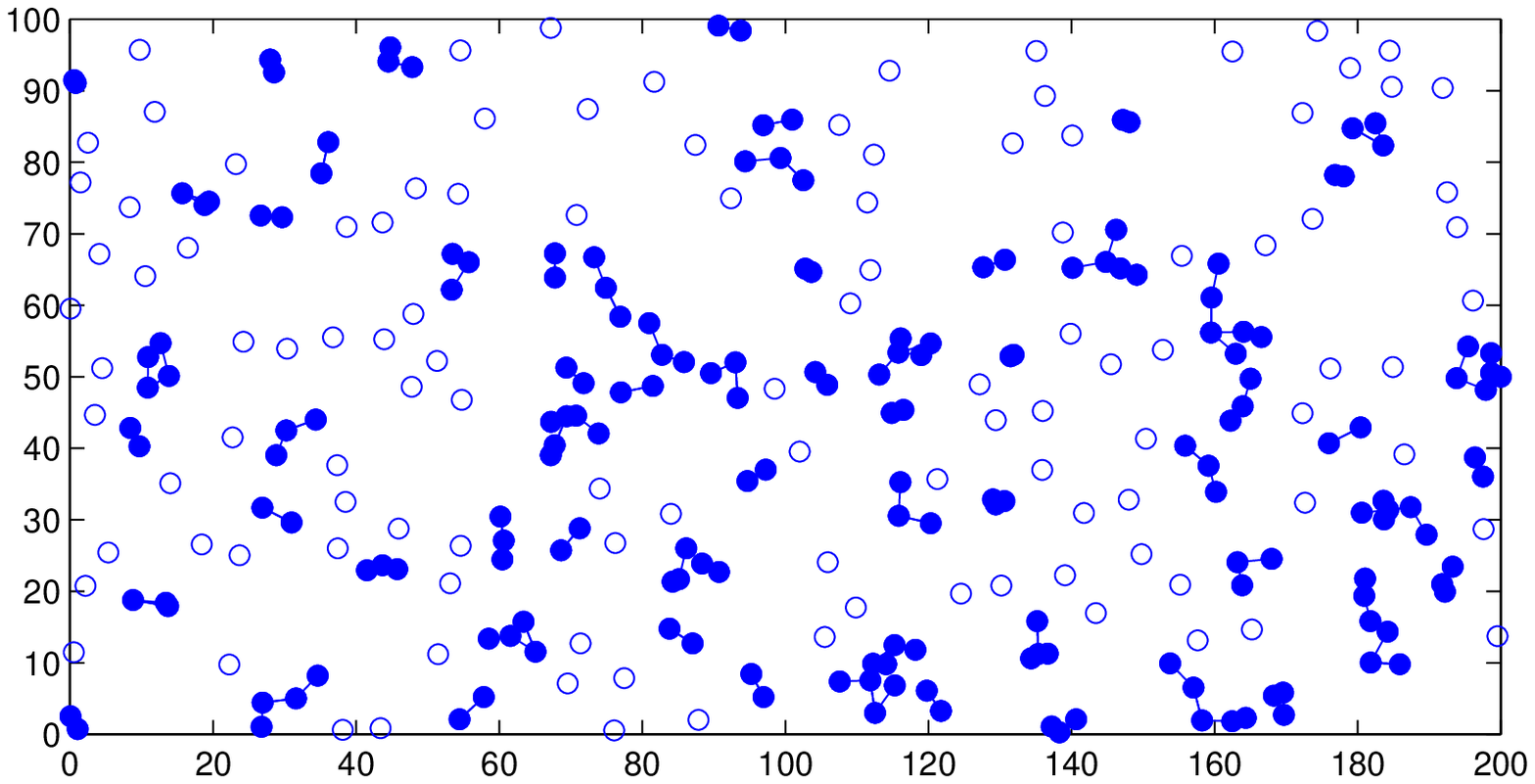, width=2.2in} } \subfigure[]{ \epsfig{file=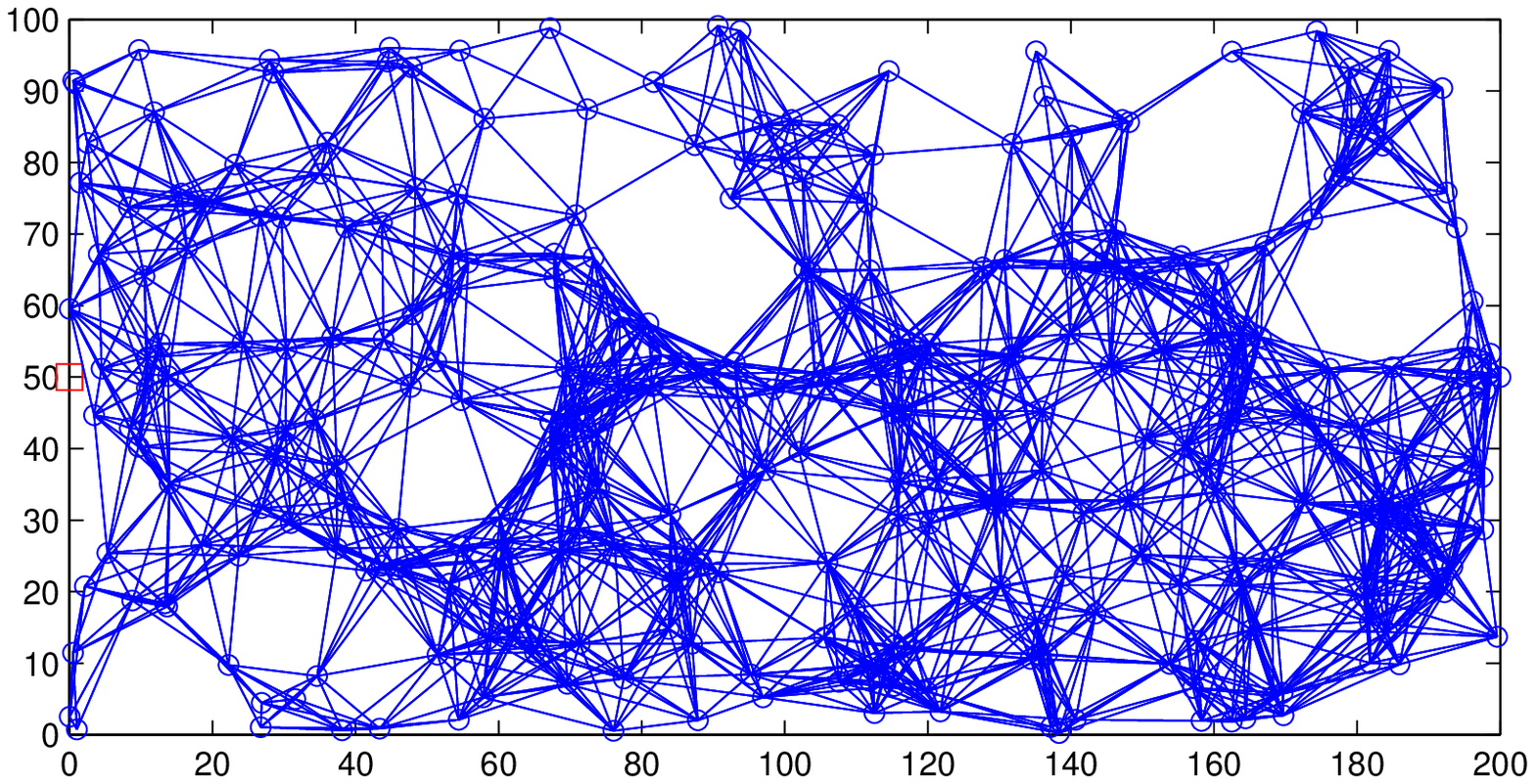, width=2.2in} }
\subfigure[]{ \epsfig{file=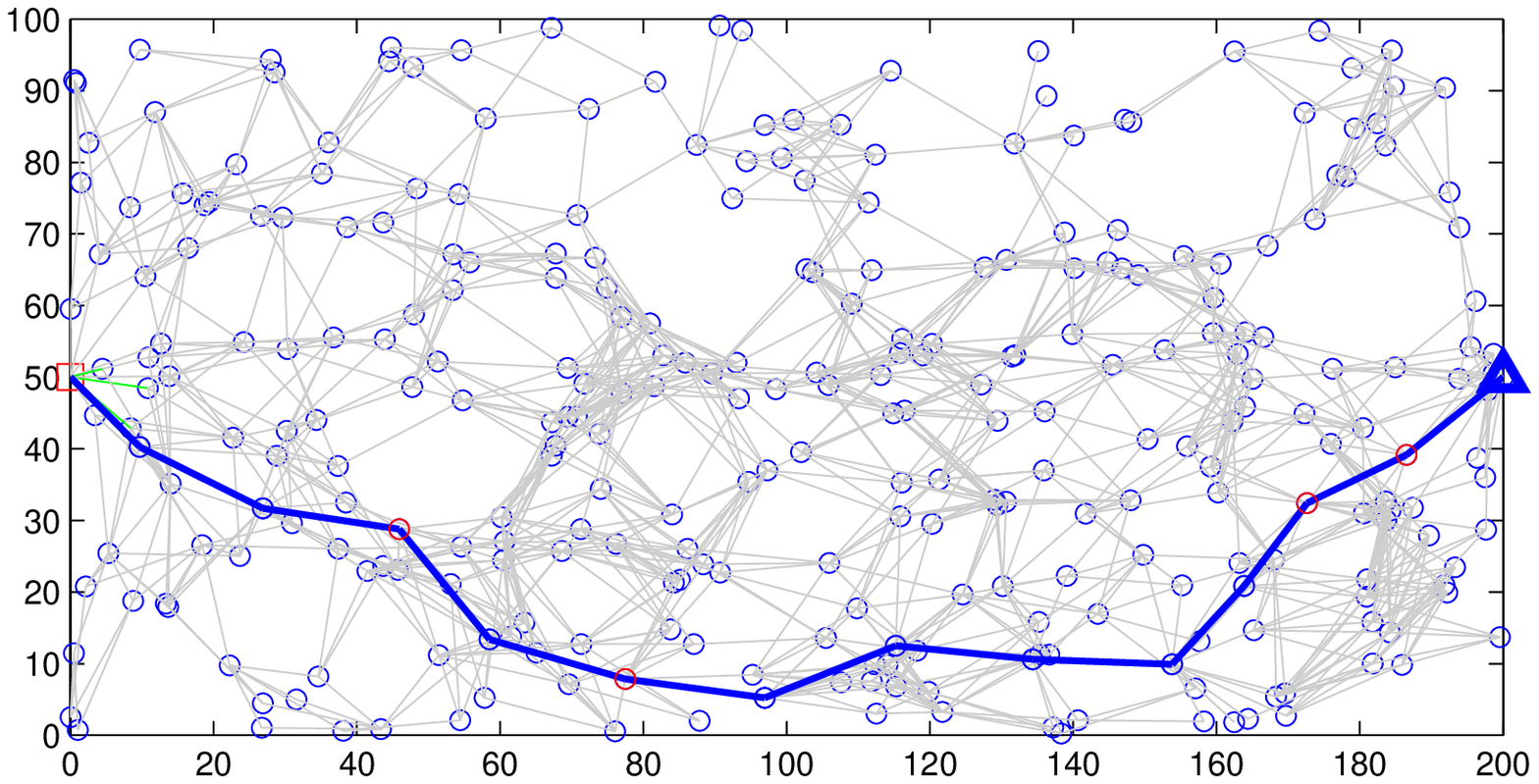, width=2.2in} } \subfigure[]{ \epsfig{file=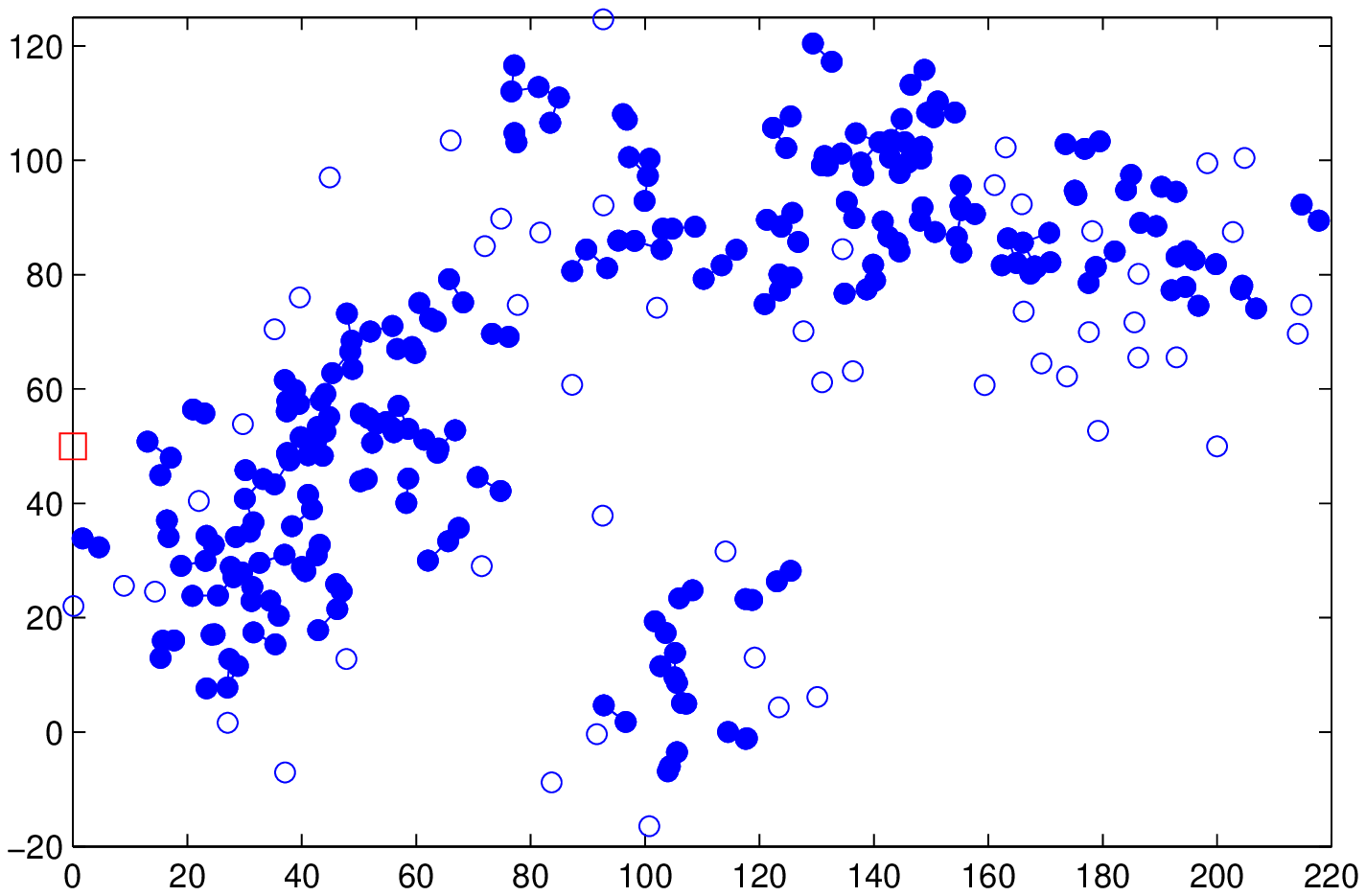, width=2.2in} }
\subfigure[]{ \epsfig{file=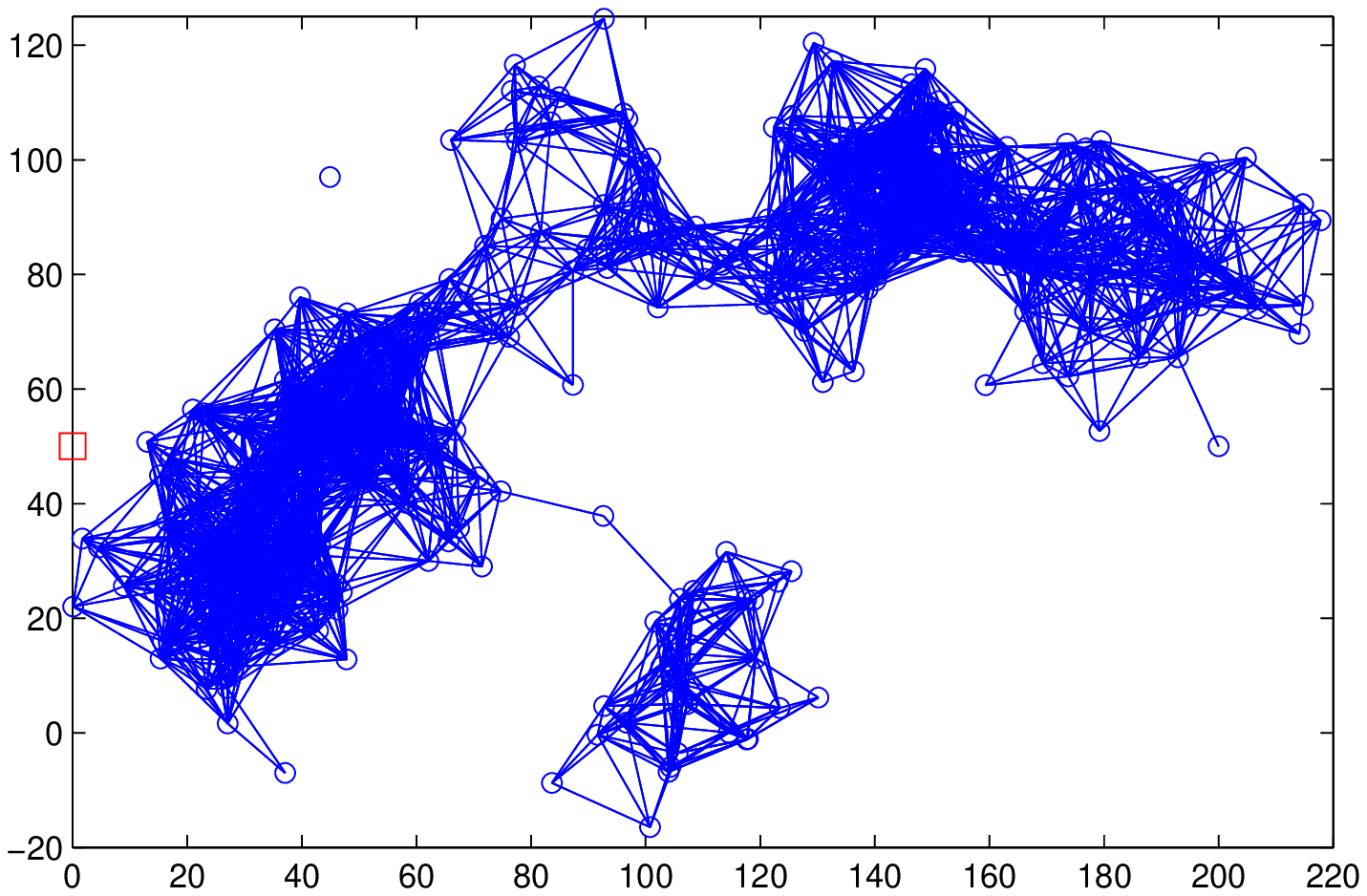, width=2.2in} } \subfigure[]{ \epsfig{file=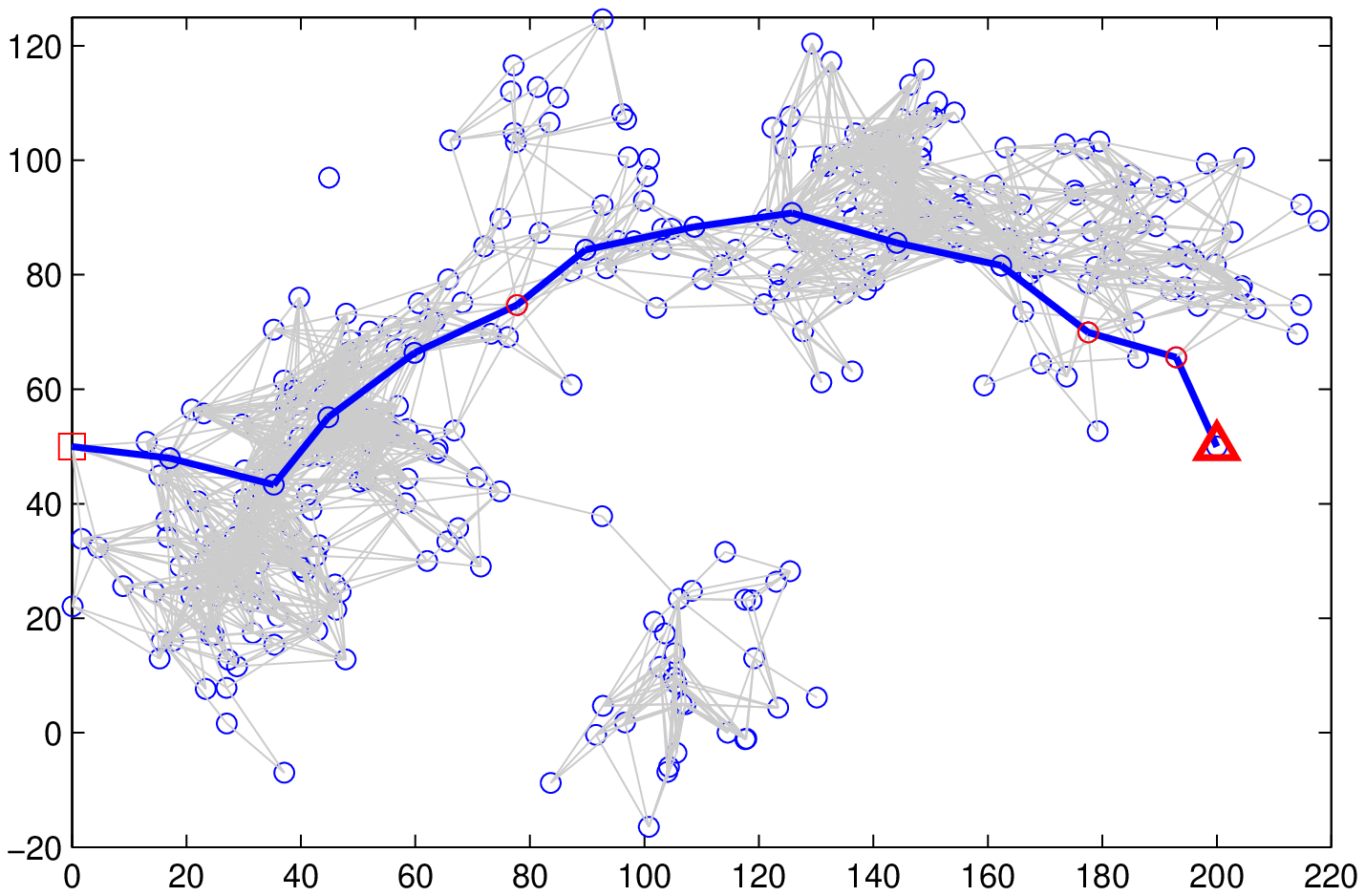, width=2.2in} }

\caption{Examples of dense wireless networks and density-based routing. (a) 300 nodes are distributed evenly in
a region of $200\times 100 \texttt{m}^2$.  If the distance between two nodes is less than $d_{max}=5$m, a
link between them is established and they are painted in blue. (b) The communication links of subfigure (a) given
the communication radius be 20m. (c) A path of density-based routing from Alice at (200,50) to the base station
at (0,50). The relays in red are unsecure relays. The links in grey are the flooding links established in the stage 1.
(d) 300 nodes are distributed unevenly in a region, and the link between nodes is established if their distance is
less than $d_{max}=5$m  and they are painted in blue. (e) The communication links of subfigure (d) given the
communication radius be 20m. (f) A path of density-based routing from Alice at (200,50) to the base station at
(0,50). The relays in red are unsecure relays. The links in grey are the flooding links established in the stage 1.}
\label{fig_random}
\end{figure*}

As illustrated in Fig. \ref{fig_random}(a) and (d), as the density increases, more nodes are close to each other
and the connected clusters (with $d_{max}=5$m) become more larger. In any connected cluster, Willie is not
able to distinguish any transmitter in it due to the lack of detection space. However if the density is not enough,
there is unconnected nodes and small clusters in the network . If we want to establish a multi-hop links to
transmit a covert message, the best way is avoiding the sparse part of the network and let network traffics flow
into the denser part of network. Based on this idea, we next propose a density-based routing algorithm to enhance
the covertness of multi-hop routing.

\begin{figure}
\centering \epsfig{file=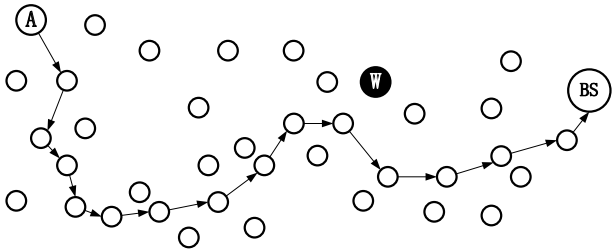, height=1.3in} \caption{Density-Based Routing (DBR).}\label{DBR}
\end{figure}

\textbf{Density-Based Routing} (DBR): DBR is designed based on the basic idea: if we can choose the relay nodes
with more neighbors, Willie will confront greater challenges to distinguish Alice from more potential transmitters.
As depicted in Fig. \ref{DBR}, Alice selects a routing path to the destination through denser node groups to avoid
being found by active Willie.

DBR is a 2-stage routing protocol which can find routes from multiple sources to a single destination, a base
station (BS). In the first stage, BS requests data by broadcasting a beacon. The beacon diffuses through the
network hop-by-hop, and is broadcasted by each node to its neighbors only once. Each node that receives the
beacon setups a backward path toward the node from which it receives the beacon.  In the second stage, a node
$i$ that has information to send to BS searches its cache for a neighboring node to relay the message. The local
rule is that, among the neighboring nodes who have broadcasted a beacon to node $i$, the node who has a larger
number of neighbors will be selected with a higher probability. Then node $i$ sends the message to the selected
relay node $j$ applied the randomized transmission scheduling. Furthermore, larger transmission probability will
be applied when node $i$ has more neighbors. Again node $j$ does the same task as node $i$ until the message
reaches BS. Algorithm 1 shows the detailed description of DBR in node $i$.

\begin{breakablealgorithm}
\caption{Density-Based Routing (Node $i$)} \hspace*{0.02in} \raggedright {\bf Input:} The set of neighbors of
node $i$: $N_{i}=\{i_1, i_2, ..., i_{k}\}$ ($i_k$ is the locally unique identifier of node), the number of neighbors
of nodes in set $N_i$: $deg(i_1)$, $deg(i_2)$, ..., $deg(i_k)$, the average number of neighbors of nodes in the
whole network $\overline{deg}$, and the upper bound of default transmission probability $0<p_{max}<1$.

\hspace*{0.02in} \raggedright {\bf Output:} The relay node $c_k\in N_i$ and the transmission probability $p_i$
of node $i$.

\hspace*{0.02in} \raggedright {\bf Initialization:} The set of candidate relay nodes $R_{i}=\{\}$.

\hspace*{0.02in} \raggedright {\bf Stage 1: Beacon Broadcasting:}
\begin{enumerate}
 \item When the node $i$ receives a beacon broadcasting by its neighbors, it checks to see if this beacon is
     rebroadcasted by itself. If not, the node broadcasts the beacon to its neighbors.

  \item Once receiving a beacon broadcasted by its neighbor $i_k$, node $i$ puts $i_k$ into the set of
      candidate relay nodes $R_{i}$.
\end{enumerate}

\hspace*{0.02in} \raggedright {\bf Stage 2: Relay Selection:}
\begin{enumerate}
\item  Suppose $R_i=\{c_1,c_2,...,c_m\}$ with $deg(c_1)\leq deg(c_2)\leq ...\leq deg(c_m)$. Node generates
    a random number $r_0$ between 0 and 1. If the random number is
\begin{equation}\label{aa1}
    r_0\in \biggl(0, \frac{deg(c_1)}{\sum_{i=1}^{m}deg(c_i)}\biggr]
  \end{equation}
  then the node $c_1$ becomes the relay node; Otherwise, if $k>1$ and
  \begin{equation}\label{aa}
    r_0\in \biggl(\frac{\sum_{i=1}^{k-1}deg(c_i)}{\sum_{i=1}^{m}deg(c_i)}, \frac{\sum_{i=1}^{k}deg(c_i)}{\sum_{i=1}^{m}deg(c_i)}\biggr]
  \end{equation}
  then the node $c_k$ becomes the relay node.

  \item Node $i$ chooses his transmission probability as follows
  \begin{equation}\label{pp}
    p_i=\frac{1}{1+e^{-x_i}}\cdot p_{max}
  \end{equation}
  where $x_i=deg(i)-\overline{deg}$.

\end{enumerate}
\label{alg_II}
\end{breakablealgorithm}

\begin{figure}
\centering \epsfig{file=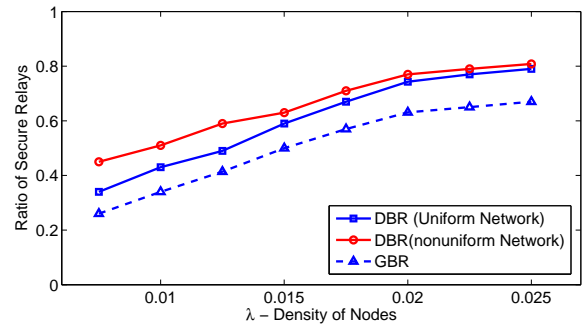, height=1.7in} \caption{The ratio of secure relays versus the density of network for different routing schemes.
The network configuration is illustrated in Fig. \ref{fig_random}. GBR is the gradient-based routing \cite{GBR}. }\label{DBR1}
\end{figure}

Fig. \ref{fig_random} illustrates examples of density-based routing when the network is uniform or nonuniform.
Fig. \ref{DBR1} depicts the ratio of secure relays versus the density of network for different routing schemes.
Two routing schemes, DBR and GBR (Gradient-Based Routing \cite{GBR}), are compared. We can find that more
dense a network is, the relays are more secure, that is, the transmitters are hidden in a noisy network, not easily
be detected by Willie. Besides, DBR has a better security performance than the GBR (Gradient-Based Routing).
Furthermore, a nonuniform network is more secure than a uniform network, since a nonuniform network has more
dense clusters that the active Willie cannot distinguish any transmitter (hidden in it) from more other potential
transmitters.

DBR is a simple routing protocol in which each node does not need sophisticated operations, such as measuring the
distance between it and its neighbors. This is reasonable since the node in an IoT network may be a very simple
node. However, if the node has the knowledge of the distance to his neighbors, the security performance may be
improved.

\section{Conclusions} \label{ch_6}
We have demonstrated that the active Willie is hard to be defeated to achieve covertness of communications. If
Alice has no prior knowledge about Willie, she cannot hide her transmission behavior in the presence of an active
Willie, and the square root law is invalid in this situation. We then propose a countermeasure to deal with the
active Willie. If Alice's transmission probability is set below a threshold, Willie cannot detect Alice's transmission
behavior for a certain significance value. We further study the covert communication in dense wireless networks,
and propose a density-based routing scheme to deal with multi-hop covert communications. We find that, as the
network becomes more and more dense and complicated,  Willie's difficulty of detection is greatly increased.
What Willie sees may become a ``shadow'' network.

As a first step of studying the effects of active Willie on covert wireless communication, this work considers a
simple scenario with one active Willie only. A natural future work is to extend the study to multi-Willie. They may
work in coordination to enhance their detection ability, and may launch other sophisticated attacks. Another
relative aspect is how to extend the results to MIMO channels. Perhaps the most difficult challenge is how to
cope with a powerful active Willie equipped with more antennas than Alice and Bob have.




\begin{thebibliography}{10}
\providecommand{\url}[1]{#1} \csname url@samestyle\endcsname \providecommand{\newblock}{\relax}
\providecommand{\bibinfo}[2]{#2} \providecommand{\BIBentrySTDinterwordspacing}{\spaceskip=0pt\relax}
\providecommand{\BIBentryALTinterwordstretchfactor}{4}
\providecommand{\BIBentryALTinterwordspacing}{\spaceskip=\fontdimen2\font plus
\BIBentryALTinterwordstretchfactor\fontdimen3\font minus
  \fontdimen4\font\relax}
\providecommand{\BIBforeignlanguage}[2]{{%
\expandafter\ifx\csname l@#1\endcsname\relax
\typeout{** WARNING: IEEEtran.bst: No hyphenation pattern has been}%
\typeout{** loaded for the language `#1'. Using the pattern for}%
\typeout{** the default language instead.}%
\else \language=\csname l@#1\endcsname \fi #2}} \providecommand{\BIBdecl}{\relax} \BIBdecl

\bibitem{Hiding_Information} B.~A. Bash, D.~Goeckel, D.~Towsley, and S.~Guha, ``Hiding information in noise:
  Fundamental limits of covert wireless communication,'' \emph{IEEE
  Communications Magazine}, 2015.

\bibitem{Computer21} L.~B. Oliveira, F.~M.~Q. Pereira, R.~Misoczki, D.~F. Aranha, F.~Borges, and
  J.~Liu, ``The computer for the 21st century: Security privacy challenges
  after 25 years,'' in \emph{2017 26th International Conference on Computer
  Communication and Networks (ICCCN)}, July 2017, pp. 1--10.

\bibitem{panda_hunter} P.~Kamat, Y.~Zhang, W.~Trappe, and C.~Ozturk, ``Enhancing source-location
  privacy in sensor network routing,'' in \emph{Proceedings of the 25th IEEE
  International Conference on Distributed Computing Systems, ICSCS'05},
  Columbus, Ohio, USA, June 2005, pp. 599--608.

\bibitem{Steganography} J.~Fridrich, \emph{Steganography in Digital Media: Principles, Algorithms, and
  Applications}, 1st~ed.\hskip 1em plus 0.5em minus 0.4em\relax Cambridge Univ.
  Press, 2009.

\bibitem{covert_channel_1} S.~Zander, G.~Armitage, and P.~Branch, ``A survey of covert channels and
  countermeasures in computer network protocols,'' \emph{IEEE Communications
  Surveys Tutorials}, vol.~9, no.~3, pp. 44--57, Third 2007.

\bibitem{covert_channel_2} S.~Cabuk, C.~E. Brodley, and C.~Shields, ``Ip covert timing channels: design
  and detection,'' in \emph{Proceedings of 11th ACM conf. Computer and
  communication security, CCS'04}.\hskip 1em plus 0.5em minus 0.4em\relax New
  York, USA: ACM, September 2004, pp. 178--187.

\bibitem{Spread_Spectrum} M.~K. Simon, J.~K. Omura, R.~A. Scholtz, and B.~K. Levitt, \emph{Spread
  Spectrum Communications Handbook}, revised edition~ed.\hskip 1em plus 0.5em
  minus 0.4em\relax McGraw-Hill, 1994.

\bibitem{square_law} B.~Bash, D.~Goeckel, and D.~Towsley, ``Limits of reliable communication with
  low probability of detection on awgn channels,'' \emph{IEEE Journal on
  Selected Areas in Communications}, vol.~31, no.~9, pp. 1921--1930, September
  2013.

\bibitem{time1} B.~A. Bash, D.~Goeckel, and D.~Towsley, ``Covert communication gains from
  adversary¡¯s ignorance of transmission time,'' \emph{IEEE Transactions on
  Wireless Communications}, vol.~15, no.~12, pp. 8394--8405, 2016.

\bibitem{SNR} R.~Tandra and A.~Sahai, ``Snr walls for signal detection,'' \emph{IEEE Journal
  of Selected Topics in Signal Processing}, vol.~2, no.~1, pp. 4--17, February
  2008.

\bibitem{LDP1} S.~Lee, R.~J. Baxley, M.~A. Weitnauer, and B.~Walkenhorst, ``Achieving
  undetectable communication,'' \emph{IEEE Journal of Selected Topics in Signal
  Processing}, vol.~9, no.~7, pp. 1195--1205, October 2015.

\bibitem{Biao_He} B.~He, S.~Yan, X.~Zhou, and V.~K.~N. Lau, ``On covert communication with noise
  uncertainty,'' \emph{IEEE Communications Letters}, vol.~21, no.~4, pp.
  941--944, April 2017.

\bibitem{Fundamental_Limits} L.~Wang, G.~W. Wornell, and L.~Zheng, ``Fundamental limits of communication
  with low probability of detection,'' \emph{IEEE Transactions on Information
  Theory}, vol.~62, no.~6, pp. 3493--3503, June 2016.

\bibitem{Bloch} M.~R. Bloch, ``Covert communication over noisy channels: A resolvability
  perspective,'' \emph{IEEE Transactions on Information Theory}, vol.~62,
  no.~5, pp. 2334--2354, May 2016.

\bibitem{Challenges} W.~Trappe, ``The challenges facing physical layer security,'' \emph{IEEE
  Communications Magazine}, vol.~53, no.~6, pp. 16--20, June 2015.

\bibitem{Artificial_noise_Goel} S.~Goel and R.~Negi, ``Guaranteeing secrecy using artificial noise,''
  \emph{IEEE Transactions on Wireless Communications}, vol.~7, no.~6, pp.
  2180--2189, June 2008.

\bibitem{heartbeats} S.~Gollakota, H.~Hassanieh, B.~Ransford, D.~Katabi, and K.~Fu, ``They can hear
  your heartbeats: Non-invasive security for implantable medical devices,'' in
  \emph{ACM SIGCOMM}, New York, NY, USA, 2011, pp. 2--13.

\bibitem{jammer1} T.~V. Sobers, B.~A. Bash, S.~Guha, D.~Towsley, and D.~Goeckel, ``Covert
  communication in the presence of an uninformed jammer,'' \emph{IEEE
  Transactions on Wireless Communications}, vol.~16, no.~9, pp. 6193--6206,
  2017.

\bibitem{jammer2} R.~Soltani, B.~Bashy, D.~Goeckel, S.~Guhaz, and D.~Towsley, ``Covert single-hop
  communication in a wireless network with distributed artificial noise
  generation,'' in \emph{Fifty-second Annual Allerton Conference}, Allerton
  House, UIUC, Illinois, USA, October 2014, pp. 1078--1085.

\bibitem{DBLP:journals/corr/abs-1709-07096} R.~Soltani, D.~Goeckel, D.~Towsley, B.~A. Bash, and S.~Guha,
    ``Covert wireless
  communication with artificial noise generation,'' \emph{CoRR}, vol.
  abs/1709.07096, 2017.

\bibitem{The_Sound_and_the_Fury} Z.~Liu, J.~Liu, Y.~Zeng, L.~Yang, and J.~Ma, ``The sound and the fury:
    Hiding
  communications in noisy wireless networks with interference uncertainty,''
  \emph{CoRR}, vol. abs/1712.05099, 2017.

\bibitem{Biao_he_2} B.~He, S.~Yan, X.~Zhou, and H.~Jafarkhani, ``Covert wireless communication with
  a poisson field of interferers,'' \emph{CoRR}, vol. abs/1712.07062, 2017.

\bibitem{cox} D.~R. Cox and A.~Stuart, ``Some quick sign tests for trend in location and
  dispersion,'' \emph{Biometrika}, vol.~42, no. 1/2, pp. 80--95, 1955.

\bibitem{radio_fingerprinting} K.~B. Rasmussen and S.~Capkun, ``Implications of radio fingerprinting on the
  security of sensor networks,'' in \emph{2007 Third International Conference
  on Security and Privacy in Communications Networks and the Workshops -
  SecureComm 2007}, Sept 2007, pp. 331--340.

\bibitem{Haenggi_PPP} M.~Haenggi, \emph{Stochastic Geometry for Wireless Networks}, 1st~ed.\hskip 1em
  plus 0.5em minus 0.4em\relax New York, NY, USA: Cambridge University Press,
  2012.

\bibitem{GBR} C.~Schurgers and M.~B. Srivastava, ``Energy efficient routing in wireless
  sensor networks,'' in \emph{2001 MILCOM Proceedings Communications for
  Network-Centric Operations: Creating the Information Force (Cat.
  No.01CH37277)}, 2001, pp. 357--361 vol.1.

\end{thebibliography}


\end{document}